\documentclass{amsart}
\usepackage{lipsum}
\pdfoutput=1
\usepackage[pdftex]{graphics}
\usepackage{amsmath,amssymb,mathtools}
\usepackage{graphicx}
\usepackage{xcolor}
\usepackage{epsdice}
\usepackage{enumitem}
\usepackage[colorlinks]{hyperref}
\usepackage[all]{xy}

\graphicspath{{./Figures/}}

\newtheorem{theorem}{Theorem}[section]

\theoremstyle{definition}

\newtheorem{conjecture}[theorem]{Conjecture}
\newtheorem{example}[theorem]{Example}

\theoremstyle{remark}

\def\CA{\mathcal{A}}

\def\a{\alpha}

\def\a{\alpha}

\begin{document}

\author{Prarit Agarwal}
\address{Queen Mary University of London\\
	Mile End Road, London E1 4NS \\ UK}
\address{Elaitra Ltd}
\email{agarwalprarit@gmail.com}

\author{Dongmin Gang}
\address{Department of Physics and Astronomy \& Center for Theoretical Physics \\
	Seoul National University \\ 1 Gwanak-ro \\ Seoul 08826 \\ Korea}
\address{Asia Pacific Center for Theoretical Physics (APCTP), Pohang 37673, Korea}
\email{arima275@snu.ac.kr}

\author{Sangmin Lee}
\address{College of Liberal Studies\\ Seoul National University\\Seoul 08826, Korea}
\address{Department of Physics and Astronomy \& Center for Theoretical Physics \\
	Seoul National University \\ 1 Gwanak-ro \\ Seoul 08826 \\ Korea}
\email{sangmin@snu.ac.kr}

\author{Mauricio Romo}
\address{Yau Mathematical Sciences Center\\ Tsinghua University\\ Beijing\\ 100084\\ China}
\email{mromoj@tsinghua.edu.cn}

\title[Quantum trace in 3-manifold and Volume conjecture]{Quantum trace map for 3-manifolds \\ and a `length conjecture'}

\begin{abstract}

We introduce a {\it quantum trace} map  for an ideally triangulated hyperbolic knot complement $S^3\backslash \mathcal{K}$. The map assigns a quantum operator to each element of Kauffmann Skein module of the 3-manifold. The quantum operator lives in a module generated by products of quantized edge parameters of the ideal triangulation modulo some equivalence relations determined by gluing equations. Combining the quantum map with a state-integral model of $SL(2,\mathbb{C})$ Chern-Simons theory, one can define  perturbative  invariants of knot $K$ in the knot complement whose leading part is determined by its complex hyperbolic length.  We then conjecture that the perturbative invariants determine an asymptotic expansion of the Jones polynomial for a link composed of $\mathcal{K}$ and $K$. We propose the explicit quantum trace map for figure-eight knot complement and confirm the {\it length conjecture} up to the second order in the asymptotic expansion both numerically and analytically.

\end{abstract}


\maketitle
\tableofcontents

\section{Introduction}
Kauffman bracket skein modules (KBSMs) were independently introduced by J. H. Przytycki \cite{przytycki2006skein} and V. G. Turaev \cite{turaev1988conway} based on the Kauffman bracket \cite{kauffman1988statistical} in an attempt to generalize  knot polynomials in $S^3$ to those in arbitrary 3-manifolds. The module becomes a non-commutative algebra when the 3-manifold is chosen to be a thickened surface $S \times I$ with  marked points on $S$. For the case, F. Bonahon and H. Wong in \cite{bonahon2010quantum} constructed an injective algebra homomorphsism, called {\it quantum trace map}, from the Skein algebra to the Chechov-Fock algebra of $S$. The Chechov-Fock algebra is a quantization of Teichm\"uller space using Thurston's shear coordinates \cite{thurston1998minimal,penner1987decorated} associated to  an ideal triangulation of $S$.  In terms of $SL(2,\mathbb{C})$ Chern-Simons theory, the quantum operators correspond to    Wilson loop operators whose trajectory is confined on the surface $S $. Classically the Wilson loops can be represented by a function on the phase space $P(S)$ associated  with the surface $S$. The phase space is  the moduli space of $SL(2,\mathbb{C})$ flat connections on $S$. Shear coordinates provides a natural coordinate of the phase space and the loop operators can be given as a Laurent polynomial of the coordinates.  After quantization, the phase space becomes a Hilbert-space on which the quantized loop operators naturally act. Via the 2D quantum trace map, one can express the  loop operators in terms of Laurent polynomial of the  quantized shear coordinates. 

Generalizing the idea of quantum trace map to a general 3-manifold $M$ seems to be pointless since the KBSM is  just a module instead of forming an algebra. Contrary to the common belief, we suggest that there is a natural unique 3D quantum trace map. 
 Wilson loop operators in the Chern-Simons theory  can be defined along arbitrary links on a 3-manifold. 
Classically, the  loop operator can be regarded as a function on $\mathcal{L}(M)$, the moduli space of $SL(2,\mathbb{C})$ flat connections on $M$. Unlike $P(S)$, there is no natural symplectic structure on $\mathcal{L}(M)$. Instead, the space $\mathcal{L}(M)$ can be regarded as a Lagrangian subvariety of the phase space $P(\mathbb{T}^2)$ when the 3-manifold has a torus  ($\mathbb{T}^2$) boundary.
 For an ideally triangulated 3-manifold $M$, the moduli space can  be represented by an algebraic variety determined by gluing equations \cite{thurston1979geometry,neumann1985volumes}. Gluing equations are set of  algebraic equations among edge parameters of ideal tetrahedra  in the triangulation. The gluing equations have a symplectic structure and the quantization of the $\mathcal{L}(M)$ has been well-studied \cite{Dimofte:2011gm,Dimofte:2012qj}. As a result of the quantization,  state-integral models \cite{hikami2007generalized,Dimofte:2011gm} are developed which  compute the partition function of $SL(2,\mathbb{C})$ Chern-Simons theory on $M$. 
 To define a 3D quantum trace map, one needs to quantize  functions on $\mathcal{L}(M)$. The functions are given by a  Laurent series of the edge parameters subjected to gluing equations. To deal with the `quantum functions' on $\mathcal{L}(M)$, we introduce  a {\it quantum gluing module} in Section \ref{sec: quantum gluing module} which are the space of them.  Our quantum trace map is an injective  module homomorphism from the KBSM of $M$ to the quantum gluing module. Once the correct quantum trace map is given, one can generalize the state-integral models with insertion of the Wilson loop operators along arbitrary knots or links in $M$. 

One big motivation for studying $SL(2,\mathbb{C})$ Chern-Simons theory on a hyperbolic knot complement $S^3\backslash \mathcal{K}$ is its relation to the volume conjecture \cite{kashaev1994quantum,kashaev1995link,murakami2001colored}. The conjecture relates a large $n$  limit of the colored Jones polynomial $J_n(\mathcal{K};q)$ to the hyperbolic volume of its knot complement, $S^3\backslash \mathcal{K}$.  As its strongest version \cite{gukov2005three}, it is conjectured that the  asymptotic limit is fully determined by a perturbative expansion of the state-integral on $M=S^3\backslash \mathcal{K}$. Our 3D quantum trace map can be naturally blended with these developments. 
We propose that the large $n$ limit of the ratio $J_{n,\tilde{n}=2}(\mathcal{K}\cup K;q)/J_{n}(\mathcal{K};q)$  can be fully captured by the state-integral model for $S^3\backslash \mathcal{K}$ with an insertion of  quantum trace operator associated with the knot $K $. We think of $\mathcal{K}$ as a heavy knot with large color $n$ while $K$ as a light knot with fixed color, $\tilde{n}=2$. The leading order result is given  by a simple function of the hyperbolic length of the light knot $K$ 
in the knot complement $S^3\backslash \mathcal{K}$, and we call it a {\it length conjecture} for an obvious reason.

\section{Quantum trace map}
After a brief review on the trace map of 3-manifold, we introduce a quantum version of the map which we call {\it quantum trace map}. 

\subsection{Kauffman Skein modules} 

We begin by recalling the definitions of Kauffman Skein modules 
from \cite{przytycki1997Skein,bullock1999understanding}. Let $\mathcal{K}$ be a hyperbolic knot in $S^3$ and $M$ be the knot complement: $M = S^3\backslash \mathcal{K}$. 
We restrict our discussion to hyperbolic knot complements. Some of the statements below should be modified when $M$ is a general 3-manifold. 

\begin{figure}[htbp]
	\begin{center}
		\includegraphics[width=.25\textwidth]{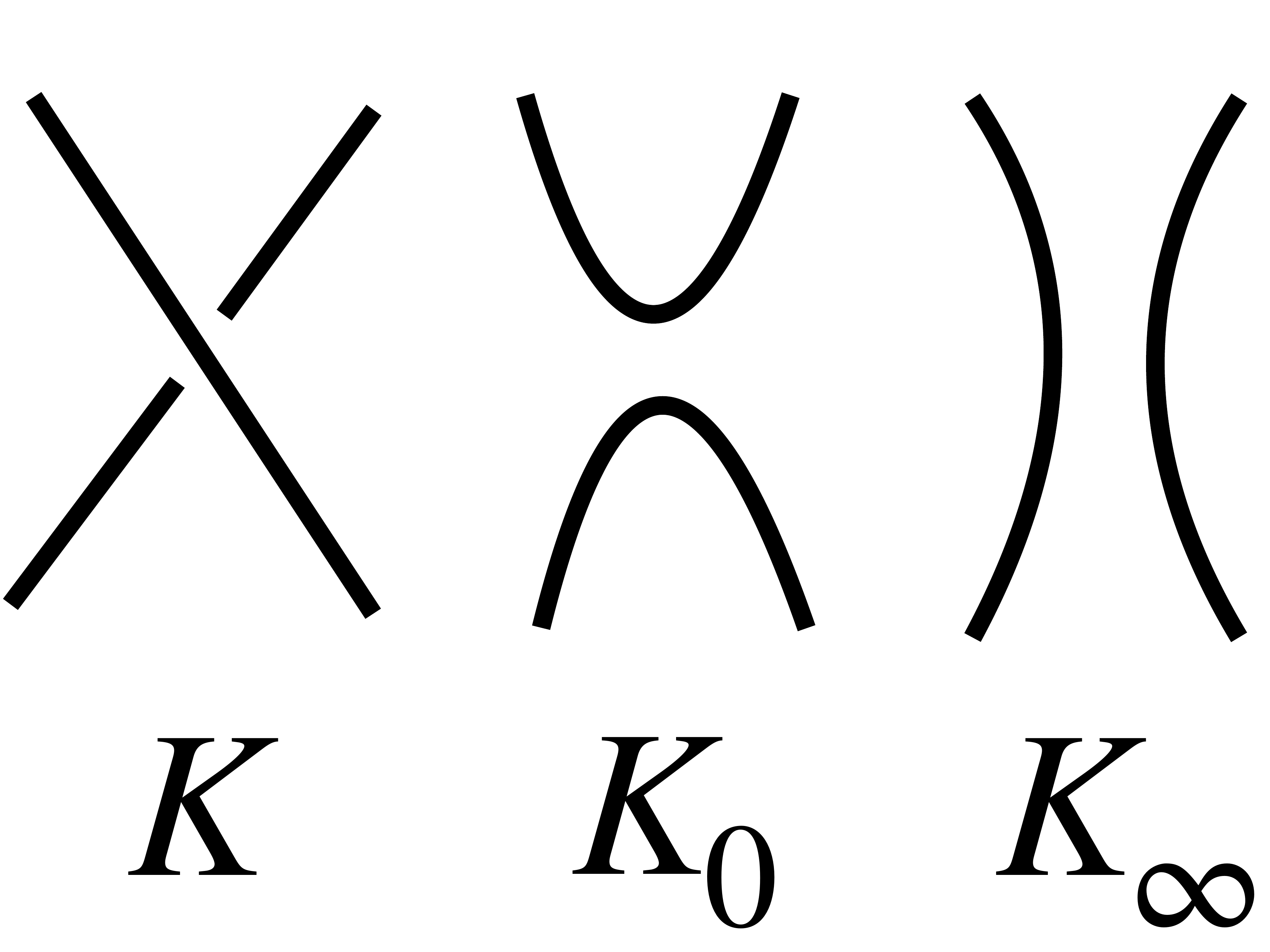}
	\end{center}
	\caption{Kauffman triple}
	\label{fig: Kauffman bracket}
\end{figure}

We denote the Kauffman bracket Skein module on $M$ by $\mathcal{S}_q[M]$ and its `even' submodule by $\mathcal{S}^{\rm even}_q[M]$. The two modules are defined as
\begin{align}
\begin{split}
\mathcal{S}_q[M] &:=\frac{\mathbb{C}[q^{\pm 1/4}]\textrm{-module with basis } \{Y_K\}_\textrm{($K$:  framed  link  in $M$)}}{\langle Y_{K_+}-q^{-1/4} Y_{K_0} -q^{1/4} Y_{K_{\infty}}, \; Y_{K \bigsqcup	\bigcirc}  + (q^{1/2} +q^{-1/2}) Y_K  \rangle},
\\
\mathcal{S}^{\rm even}_q[M] &:= \frac{\mathbb{C}[q^{\pm 1/4}]\textrm{-module with basis } \{Y_K\}_\textrm{($K$:  framed `even' link  in $M$)}}{\langle Y_{K_+}-q^{-1/4} Y_{K_0} -q^{1/4} Y_{K_{\infty}}, \; Y_{K \bigsqcup	\bigcirc}  + (q^{1/2} +q^{-1/2}) Y_K  \rangle}\;.
\end{split} \label{Sq[M]}
\end{align}
The basis includes $Y_{K=\emptyset}$, where $\emptyset$ is the empty link.  Here, $K_+,K_0$ and $K_{\infty}$, are three framed links which are identical except in a small 3-ball, 
        as depicted in Figure \ref{fig: Kauffman bracket}. 
        $K \bigsqcup \bigcirc$ is the union of $K$ with an unlinked, 0-framed unknot in the trivial homotopy class. 
The $Y_K$ in $\mathcal{S}_q[M]$ is invariant under Reidemeister moves II, III and modified Reidemeister move I on $K$ but not under Reidemeister move I. 
\begin{figure}[htbp]
	\begin{center}
		\includegraphics[width=.25\textwidth]{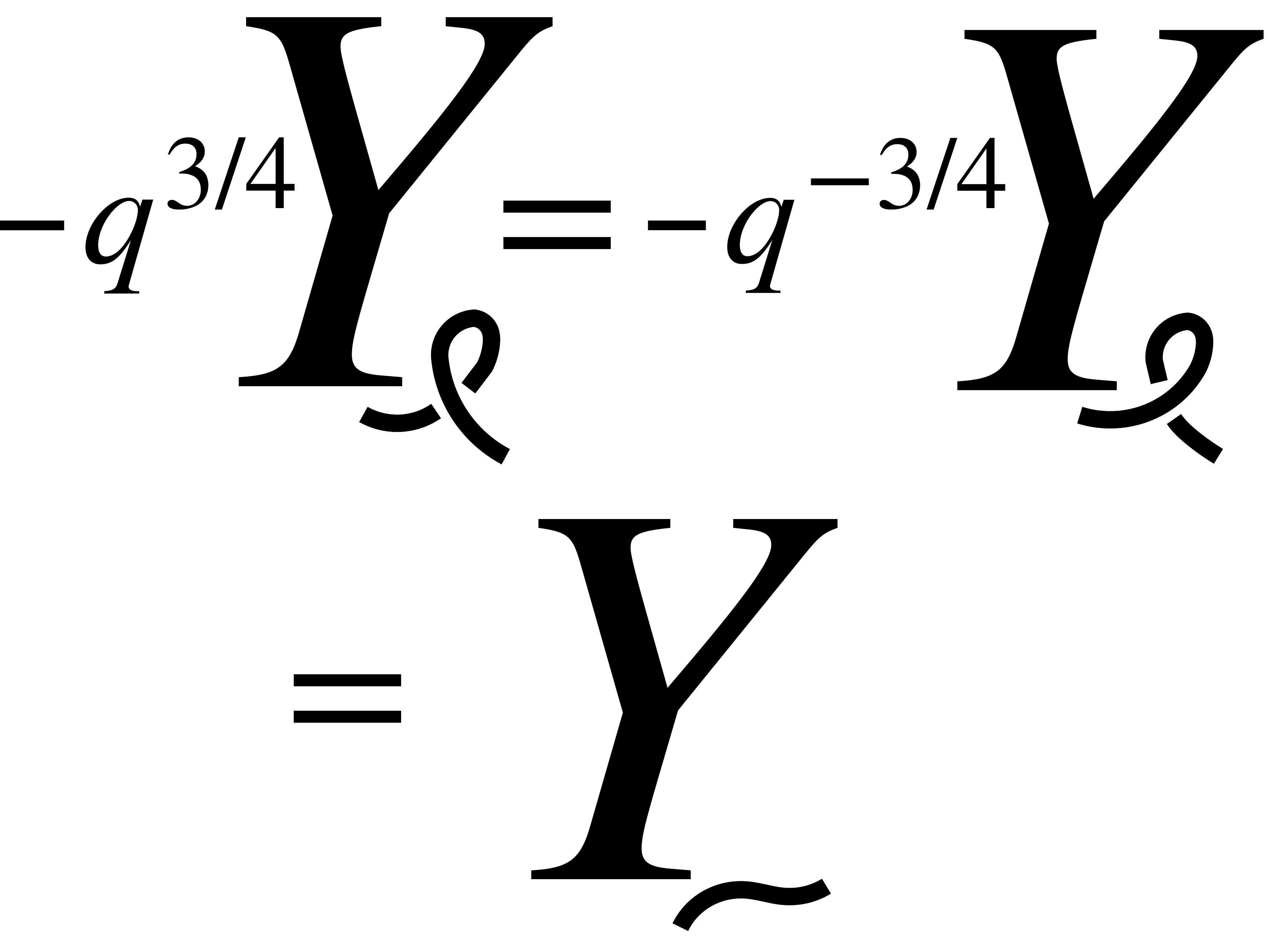}
	\end{center}
	\caption{$Y_K$ under Reidemeister move I on $K$.}
	\label{fig: Reidemeister move 1}
\end{figure}
This is why the $Y_K$ is labelled by {\it framed} link $K$.
We call a link $K = \bigcup_{I=1}^{\sharp(K)} K_I$ consisting of $\sharp(K)$ knots  an `even'  link 
if 
\[
\prod_{I=1}^{\sharp(K)} \textrm{sgn} (K_I)=+1 \;,
\] 
and an `odd' link otherwise. Here $\textrm{sgn} (K_I) \in \{ +1 , -1 \}$ is defined as
\begin{align}
\textrm{sgn} (K_I) := 
\begin{cases}
+1, \quad \textrm{if $[K_I]=1$ as an element of $H_1 (M,\mathbb{Z}_2) = \mathbb{Z}_2 =\{+1,-1 \}$\;, }\\
-1, \quad \textrm{otherwise.}
\end{cases}
\end{align}
For later use, we consider a submodule $\mathcal{S}_q^{\rm even}[M]$ whose basis $Y_K$ are labelled by even links $K$ on $M$. Note that the Kauffman bracket Skein relation is  well-defined in $\mathcal{S}_q^{\rm even}$, i.e. all three links in Kauffman triple  share the same evenness/oddness and $K \bigsqcup \bigcirc$ is even if $K$ is even.

\subsubsection{Skein algebra and trace map} 

In the special case $q^{1/2}=1$, the Skein module gives rise to the {\it Skein algebra}, 
$V[M]:=\mathcal{S}_q[M]|_{q^{1/2}=1}$.

\begin{figure}[htbp]
	\begin{center}
		\includegraphics[width=.15\textwidth]{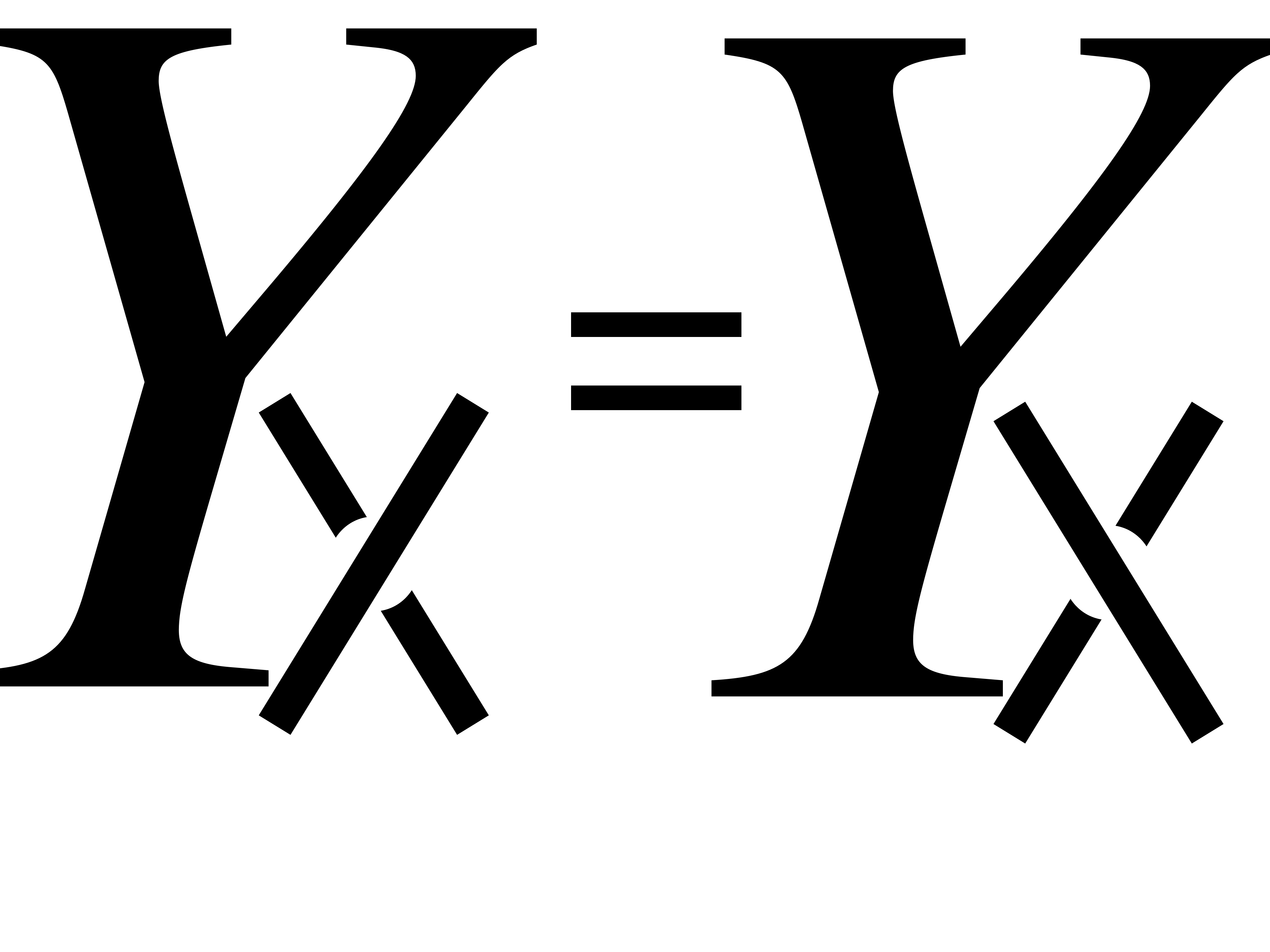}
	\end{center}
	\caption{Kauffman bracket Skein relation at $q^{1/2}=1$.}
	\label{fig: Kauffman bracket at q=1}
\end{figure}
At $q^{1/2}=1$, the Skein relation equates the over-crossing and the under-crossing, 
as in Figure~\ref{fig: Kauffman bracket at q=1}, 
so that the Skein algebra is a indeed a commutative $\mathbb{C}$-algebra 
equipped with the multiplication,  
\begin{align}
Y_{K_1} Y_{K_2}  =Y_{K_2} Y_{K_1} := Y_{K_1 \cup K_2} \;. 
\end{align}   
The $\mathbb{C}$-algebra  is generated by elements in  $\pi_1 (M)$ modulo some relations: 
\begin{align}
\begin{split}
V[M] &=\frac{ \mathbb{C}[Y_\gamma : \gamma \in \pi_1(M) ]}{ \langle Y_{\gamma_1} Y_{\gamma_2}-Y_{\gamma_2}Y_{\gamma_1},\;Y_{\gamma_1} Y_ {\gamma_2} +Y_{\gamma_1  \gamma_2} + Y_{\gamma_1  \gamma_2^{-1}} \;, \; Y_e +2\rangle} \;,
\\
V^{\rm even}[M] &=\{x \in V[M]\;:\; \textrm{$x$ is even}\} \;.
\end{split}
\end{align}
Here $e \in \pi_1(M)$ is the identity element.  The above relations imply that $Y_\gamma = Y_{\gamma^{-1}}$ and $Y_\gamma$  depends only on the  conjugacy class of $\gamma$ in $\pi_1 (M)$. General element $x \in V^{\rm even}[M]$ is given as 
\begin{align}
&\sum_I C_I\prod_{i=1}^{n_I} Y_{\gamma^{I}_i}\quad  \textrm{with $\gamma_i^I \in \pi_1 (M),\;C_I \in \mathbb{C}$  and $\prod_{i=1}^{n_I} \textrm{sgn}[\gamma_i^I]=1$ for all $I$.} \nonumber
\end{align}
The relations in the denominator come from the relations  in the Skein module $\mathcal{S}_q[M]$ in eqn.\eqref{Sq[M]} at $q^{1/4}=-1$.  We choose $q^{1/4}=-1$ instead of $q^{1/4}=1$ for a later convenience. There is an algebra isomorphism between $\mathcal{S}_q[M]$ at $q^{1/4}=1$ and $\mathcal{S}_q[M]$ at $q^{1/4}=-1$.

One interesting aspect of the Skein algebra is its relation to the coordinate ring of $SL(2,\mathbb{C})$ character variety $X[M]$, which is defined as 
\begin{align}
\begin{split}
&X[M] := \textrm{Hom}\big{[}\pi_1 (M)\rightarrow SL(2,\mathbb{C}) \big{]}/\sim 
\\
& \textrm{where }\rho_1 \sim \rho_2 \textrm{ if } \textrm{Tr}(\rho_1 (\gamma)) = \textrm{Tr} (\rho_2 (\gamma)) \textrm{ for all $\gamma \in \pi_1(M)$}.
\end{split}
\end{align}
For our purposes, we consider the $\mathbb{Z}_2$-quotient of the variety,  $X[M]/\mathbb{Z}_2$, where the $\mathbb{Z}_2$ is the action of $H^1(M, \mathbb{Z}_2) = \textrm{Hom}(\pi_1 M \rightarrow \mathbb{Z}_2 =\{\pm 1\})= \{1, \eta\}$:
\begin{align}
	\begin{split}
&X[M]/\mathbb{Z}_2  =\{ \rho \in X[M]\}/\sim\;
	\\
&\textrm{where } \rho_1 \sim \rho_2 \textrm{ if } \rho_1 (\gamma) = \eta (\gamma)\rho_2 (\gamma)\; \textrm{ for all }\gamma\in \pi_1 (M)\;.
\end{split}
\end{align}
One can naturally define an algebra homomorphism $\Phi_{\rm tr}$ called {\it trace map} 
from the Skein algebra $V[M]$ to the coordinate ring $\mathbb{C}[X(M)]$ 
(or from $V^{\rm even}[M]$ to $\mathbb{C}[X(M)/\mathbb{Z}_2]$), 
the algebra of functions on $X(M)$, as follows
\begin{align}
\begin{split}
&\Phi_{\rm tr} \; :\; V(M) \;\left(\textrm{or } V^{\rm even}(M) \right)\rightarrow \mathbb{C}[X(M)]\; \left(\textrm{or }\mathbb{C}[X(M)/\mathbb{Z}_2]\right), \textrm{ where }
\\
& [ \textrm{$\Phi_{\rm tr}$}(Y_\gamma)] (\rho) = -  \textrm{Tr} (\rho (\gamma))\;.
\end{split} \label{classical trace map}
\end{align}
To see that the map is an algebra homomorphism, one needs to use the following property of $SL(2,\mathbb{C})$ matrices
\begin{align}
\textrm{Tr}(A) \textrm{Tr}(B) = \textrm{Tr}(AB) + \textrm{Tr}(AB^{-1})\;, \quad \textrm{Tr}(\mathbb{I})=2\;,
\end{align}
where $\mathbb{I}$ is the $2\times 2$ identity matrix. 
The trace map is well-defined as a homomorphism from $V^{\rm even}(M)$ to $\mathbb{C}[X(M)/\mathbb{Z}_2]$ since $\prod_{I=1}^n \eta (\gamma_I)=1$ for  $\prod_{I=1}^{n} Y_{\gamma_I} \in V^{\rm even}(M)$.  
This section can be summarized by the diagram in Figure~\ref{fig: algebra and map}. 
\begin{figure}[htbp]
	\[\xymatrix{
	\qquad    \mathcal{S}^{\rm even}_q[M] \ar[d]^{q^{1/2}=1} &  \\
		\Phi_{\rm tr}\;:\; V^{\rm even}[M]\;\; \;\;\ar[r]&  \;\;\;\mathbb{C}[X(M)/\mathbb{Z}_2]
	}\]
	\caption{Relations among $\mathcal{S}^{\rm even}_q[M]$ (even Kauffman bracket Skein module), $V^{\rm even}[M]$ (even Skein algebra) and $\mathbb{C}[X(M)/\mathbb{Z}_2]$ (coordinate ring of character variety).  The relations suggest that $\mathcal{S}^{\rm even}_q[M]$ can be considered as a quantization of $\mathbb{C}[X(M)/\mathbb{Z}_2]$.}
	\label{fig: algebra and map}
\end{figure}

For later use, we  define the canonical component $X_0[M] \subset X[M]$ for a hyperbolic knot complement $M=S^3\backslash \mathcal{K}$ as 
\begin{align}
	\begin{split}
		X_0[M] &:= \mbox{a connected component of $X[M]$ which contain $\rho_{\rm hyp}$}\;,
		\\
		\rho_{\rm hyp} &:= \mbox{an $SL(2,\mathbb{C})$-representation} 
		\\
		&\quad \quad \mbox{corresponding to the complete hyperbolic structure}\;. \label{X0 variety}
	\end{split}
\end{align}
In the same way, one can define $X_0(M)/\mathbb{Z}_2$.

\subsection{Quantum gluing module} \label{sec: quantum gluing module} Following \cite{Dimofte:2011gm,Gang:2015wya}, 
we define the quantum gluing module $\hat{\mathbb{C}}_q [\mathcal{T}]$ associated to an ideal triangulation  $\mathcal{T}$ of $M$ as
\begin{align}
\begin{split}
&\hat{\mathbb{C}}_q [\mathcal{T}] =\frac{ \left( \mathbb{C}[q^{\pm 1/4}][\hat{z}_i, \hat{z}''_i, \hat{z}^{-1}_i , \hat{z}_i''^{-1}])\big{/}\langle \hat{z}_i \hat{z}''_j - q^{-\delta_{ij}} \hat{z}''_j \hat{z}_i \rangle \right)}{\textrm{equivalence relations}\sim}
\\
&\textrm{where }\hat{O} \sim e^{\hat{C}_I} \cdot\hat{O}\;\textrm{and}\;  \hat{O} \sim \hat{O}\cdot (\hat{z}^{-1}_i + \hat{z}''_i)\;.
\end{split} \label{quantum gluing module}
\end{align}
Here $i$ runs from 1 to $|\mathcal{T}|$, the number of tetrahedra in the ideal triangulation.
The precise definition of the exponentiated internal edge operators $e^{\hat{C}_I}$ can be found in \eqref{exponentiated C} 
in appendix~\ref{gluing}. 
Before taking the quotient by the equivalence relations, it is a non-commutative algebra. 
After the quotient, it becomes just a $\mathbb{C}[q^{\pm 1/4}]$-module 
since the non-commutative multiplication is not compatible with the equivalence relation 
and thus can not be well-defined in $\hat{\mathbb{C}}_q[\mathcal{T}]$.

At $q^{1/2}= 1$, $\hat{\mathbb{C}}_q [\mathcal{T}]$ becomes the algebra of functions on the gluing equation variety $\chi[\mathcal{T}]$ 
\begin{align}
\begin{split}
& \hat{\mathbb{C}}_q[\mathcal{T}]  |_{q^{1/2}=1} \simeq  \mathbb{C}[\chi[\mathcal{T}]]
\;,
\\
&\chi[\mathcal{T}] := \{ z_i, z_i''\;:\; z_i^{-1}+z_i''-1=0, \;e^{C_I}=1\}_{i=1}^{|\mathcal{T}|}\;.
\end{split}
\end{align}
For each element $(z_i, z_i'') \in \chi[\mathcal{T}]$, there is an associated representation \cite{thurston1979geometry}
\begin{align}
\rho^{\mathcal{T}}_{(z_, z'')} \in \textrm{Hom} [\pi_1 (S^3\backslash \mathcal{K} )\rightarrow PSL(2,\mathbb{C}) ]
\end{align}
We choose an ideal triangulation $\mathcal{T}$ which is a $\rho_{\rm hyp}$-regular:
\begin{align}
\textrm{$\mathcal{T}$ is called $\rho_{\rm hyp}$-regular if $\exists (z,z'')_{\rm hyp} \in \chi[\mathcal{T}]$  such that $\rho^{\mathcal{T}}_{(z,z'')_{\rm hyp}}  = \rho_{\rm hyp}$}\;.
\end{align}
$\rho_{\rm hyp}$ is an $SL(2,\mathbb{C})$-representation in \eqref{X0 variety} but also can be regarded as an $PSL(2,\mathbb{C}) = SL(2,\mathbb{C})/\mathbb{Z}_2$-representation.  Let us denote by $\chi_0 [\mathcal{T}]$ the connected component of $\chi[\mathcal{T}]$ which contains $(z,z'')_{\rm hyp}$.
In general, the gluing equation variety  $\chi[\mathcal{T}]$ depends on the choice of $\rho_{\rm hyp}$-regular triangulation $\mathcal{T}$  while the $\chi_0[\mathcal{T}]$ does not \cite{tillmann2012degenerations}. 
The component can be identified with 
\begin{align}
\chi_0 [\mathcal{T}]\simeq X_0 (M)/\mathbb{Z}_2\;.
\end{align}
Another nice reason for restricting to the component  $\chi_0[\mathcal{T}]$ is that any $PSL(2,\mathbb{C}) = SL(2,\mathbb{C})/\mathbb{Z}_2$ representation $\rho^{\mathcal{T}}_{(z,z'')}$ with $(z,z'')\in \chi_0[\mathcal{T}]$ can be  lifted to a $SL(2,\mathbb{C})$ representation. Thus, one can define the classical trace map $\Phi_{\rm tr} : V^{\rm even}[M] \rightarrow \mathbb{C}[\chi_0[\mathcal{T}]]$ as follows
\begin{align}
[\Phi_{\rm }(Y_\gamma)](z,z'') = - \textrm{Tr}(\rho_{(z,z'')}^{\mathcal{T}}(\gamma))\;. \label{classical trace map-2}
\end{align}
The trace map is identical to the map in \eqref{classical trace map} under the identification $\chi_0(\mathcal{T})\simeq X_0[M]/\mathbb{Z}_2$. The lifting from $PSL(2,\mathbb{C})$-representation to $SL(2,\mathbb{C})$-representation  is not unique and the two different upliftings, $\rho$ and $\tilde{\rho}$, are related by the action of $H^1 (S^3\backslash \mathcal{K},\mathbb{Z}_2) = \textrm{Hom}[\pi_1 (S^3\backslash \mathcal{K}) \rightarrow \mathbb{Z}_2] =\{ 1, \eta \}$, i.e.
\begin{align}
\tilde{\rho}(\gamma) = \eta (\gamma) \rho(\gamma)\;, \quad  
\end{align}
For even $Y_\gamma \in V^{\rm even}[M]$, the trace map is well-defined, i.e. independent of the choice of liftings, since $\eta (\gamma)=1$.

\subsection{Quantum trace map and Length conjecture} 

The quantum trace map for 2D surfaces with $SL(2)$ was introduced in \cite{bonahon2010quantum}, and was generalized to groups of higher rank in \cite{Gabella:2016zxu}. Here we introduce a 3D version of the quantum trace map. 
\begin{figure}[htbp]
\[\xymatrix{
	\textrm{tr}^{\mathcal{T}}_q \;:\;& \mathcal{S}_q^{\rm even}[M] \ar[d]^{q^{1/4}=-1}\ar[r]  &  \hat{\mathbb{C}}_q[\mathcal{T}] \ar[d]^{q^{1/4}=-1} \\
	\Phi_{\rm tr}\;:\;& V^{\rm even} [M] \ar[r]&  \mathbb{C}[X_0[M]/\mathbb{Z}_2] \simeq \mathbb{C}[\chi_0[\mathcal{T}]]
	}\]
\caption{Quantum trace map.} 
\label{fig: Quantum trace map}
\end{figure}

\conjecture[main conjecture]{
There exists a unique injective module homomorphism $\textrm{tr}^{\mathcal{T}}_q : \mathcal{S}^{\rm even}_q[M]\rightarrow \hat{\mathbb{C}}_q [\mathcal{T}]$ which satisfies the following properties:} \label{main conjecture}

\vskip 0.3cm

\noindent
\textbf{I}. At $q^{1/4}=-1$, the quantum trace map is identical to  the trace map $\Phi_{\rm tr}$ 
\begin{align}
(\textrm{tr}_q^{\mathcal{T}})\big{|}_{q^{1/4}=-1} = \Phi_{\rm tr} \quad \textrm{as a map $V^{\rm even}(M) \rightarrow \mathbb{C}[X_0(M)/\mathbb{Z}_2]$} \;.
\label{conjecture 1}
\end{align}
The relation between $\textrm{tr}^{\mathcal{T}}_q$ and $\Phi_\textrm{tr}$ is summarized in Figure~\ref{fig: Quantum trace map}.

\vskip 0.3cm

\noindent
\textbf{II. (All-order length conjecture)} For an even link $K = \bigcup_{I=1}^{\sharp(K)}K_I \subset M = S^3\backslash \mathcal{K}$,
\begin{align}
\begin{split}
&  \frac{J_{n,\tilde{n}_I=2} (\mathcal{K}\cup K;q)}{J^{\rm }_{n} (\mathcal{K};q)}\bigg{|}_{q^{1/4} =- \exp(\frac{\pi i}{2k} )} \xrightarrow{ \quad n=k; \; k\rightarrow \infty \quad }  \sum_{s=0}^\infty  \mathcal{Z}^{\rm (hyp)}_s  (\hat{O}_{K};M) \left(\frac{2\pi i }k\right)^{s}\;.
\\
&\textrm{or equivalently},
\\
&\log\frac{ J_{n,\tilde{n}_I=2} (\mathcal{K}\cup K;q)}{J_{n} (\mathcal{K};q)}\bigg{|}_{q^{1/4} = -\exp(\frac{\pi i}{2k} )} \xrightarrow{ \quad n=k; \; k\rightarrow \infty \quad } \sum_{s=0}^\infty \mathcal{W}^{\rm (hyp)}_s (\hat{O}_{K};M) \left(\frac{2\pi i }k\right)^{s} \;.
\label{conjecture 2}
\end{split}
\end{align}
Here $J$ denotes a variation of the colored Jones polynomial for framed links. While the conventional colored Jones polynomial is an invariant of oriented links, our $J$ is an invariant of unoriented framed links. See Appendix \ref{Appendix : Jones polynomial} for  the definition. In the above, $I=1,\ldots ,\sharp(K)$ and $\tilde{n}_I$ denotes the `color' of the $I$-th component.

The perturbative invariant $\mathcal{W}^{\rm (hyp)}_s (\hat{O}_{K};M) $ is determined by the {\it quantum trace operator} 
 \begin{align}
 &\hat{O}_{K}:=\textrm{tr}_q^{\mathcal{T}}(Y_K) \;. 
\end{align}
More generally, for an element $Y = \sum_a C_a  Y_{K_a}  \in \mathcal{S}_q^{\rm even}[M]$, 
 \begin{align}
 &\hat{O}_{Y} := \textrm{tr}^{\mathcal{T}}_q (Y) =  \sum_a C_a \hat{O}_{K_a}\;.
 \end{align}
 The operator $\hat{O}_{K}$ is a $\hat{\mathbb{C}}_q[\mathcal{T}]$-valued invariant of unoriented framed links.
 
  By incorporating the $\hat{O}$ with the state-integral model \cite{Dimofte:2011gm},  the perturbative invariants $\{\mathcal{W}^{\rm (hyp)}_s (\hat{O}_K;M) , \mathcal{Z}^{\rm (hyp)}_s (\hat{O}_K;M)\}_{s=0}^\infty$ can be obtained from the perturbative expansion of the state-integral model. The explicit form of the state-integral model and its perturbative expansion using Feynman diagram will be given in section \ref{sec : perturbative invariants}. 
  
 Two perturbative expansions are  related to each other by exponential or logarithm, i.e.
  \begin{align}
  \exp \left(\sum_{s=0}^\infty \mathcal{W}^{\rm (hyp)}_s \hbar^{s} \right) = \sum_{s=0}^\infty \mathcal{Z}^{\rm (hyp)}_s \hbar^s \textrm{ as a formal power series in $\hbar$}. \label{ZandW}
  \end{align}
The leading ``classical" coefficient $\mathcal{Z}^{\rm hyp}_{s=0} (\hat{O}_K;M)$ is simply  given by 
\begin{align}
\mathcal{Z}^{\rm (hyp)}_{s=0} (\hat{O};M) := \left(\hat{O}|_{q^{1/4}\rightarrow -1}\right)\bigg{|}_{(z_i, z''_i) = (z'_i,z''_i)_{\rm hyp}}\;.
\end{align}
Here $\hat{O}|_{q^{1/4}\rightarrow -1}$ is regarded as an element in $\mathbb{C}[\chi_0(\mathcal{T})]$. The classical value is
  \begin{align}
  \begin{split}
& \mathcal{Z}^{\rm hyp}_{s=0} (\hat{O}_K;M)  = \textrm{Tr} \left([ \Phi_{\rm tr} (Y_{\gamma_K})] (\rho_{\rm hyp})  \right)
\;,
  \\
  &\textrm{or equivalently, }   \mathcal{W}^{\rm (hyp)}_{s=0} (\hat{O}_K;M) 
  = \log\left( [\Phi_{\rm tr} (Y_{\gamma_K})] (\rho_{\rm hyp}) \right) \,. \label{classical part : length}
  \end{split}
  \end{align}
  Here $Y_{\gamma_K} = \prod_{I=1}^n Y_ {\gamma_{I}}$ is an element of the Skein algebra obtained from $Y_K$ at $q^{-1/4}=-1$   and $\rho_{\rm hyp} \in \textrm{Hom}[\pi_1(M)\rightarrow SL(2,\mathbb{C})]$ is an $SL(2,\mathbb{C})$ flat connection corresponding to the complete hyperbolic structure on $M$. 
  The classical part is related to the complex length of the geodesic (hence the name ``length conjecture") by
  \begin{align}
   [\Phi_{\rm tr} (Y_{\gamma_K})] (\rho_{\rm hyp}) :=\prod_{I=1}^{\sharp(K)} \left(-\textrm{Tr}( \rho_{\rm hyp}(\gamma_{I}) )\right) = \prod_{I=1}^{\sharp(K)} (-e^{\frac{1}2\ell_{\mathbb{C}}(\gamma_I)}-e^{-\frac{1}2\ell_{\mathbb{C}}(\gamma_I)})\;.
  \end{align}
$\ell_{\mathbb{C}}(\gamma_I)$ denotes the complexified length of a geodesic in the same homotopy class as  $\gamma_I$ of $K_I$
  
  To prove the  length conjecture in \eqref{conjecture 2} and \eqref{classical part : length}, we only need to prove them for basis $\{Y_{K_a}\}$ of $\mathcal{S}^{\rm even}_q[M]$. It is obvious for  \eqref{conjecture 2} since for $Y_K = \sum_a C_a(q^{1/4}) Y_{K_a}$, 
  \begin{align}
 J_{n,\tilde{n}_I=2} (\mathcal{K}\cup K;q) = \sum_a C_a  (q^{1/4})J_{n,\tilde{n}_I=2} (\mathcal{K}\cup K_a;q) \,,
  \end{align}
and the perturbative invariants $\mathcal{Z}_s^{\rm (hyp)}(\hat{O}_{Y_K})$ satisfy
  \begin{align}
 	\sum_{s=0}^\infty \mathcal{Z}^{\rm (hyp)}_s (\hat{O}_{K};M) \hbar^s=  \sum_a C_a (q^{1/4} =-e^{\frac{\hbar}4}) \left( \sum_s   \mathcal{Z}^{\rm (hyp)}_s (\hat{O}_{K_a};M) \hbar^s \right)\;.
 \end{align}
The relation in  \eqref{classical part : length} is valid for arbitrary $\hat{O}_K$ if it holds for every basis since
 \begin{align}
 \begin{split}
&  \Phi_{\rm tr} (Y_{\gamma_K}) (\rho) = \sum_{a } C_a (q^{1/4}=-1) \Phi_{\rm tr} (Y_{\gamma_K}) (\rho) \;, 
\\
&\textrm{for arbitrary $\rho\in \textrm{Hom}\left[\pi_1 M \rightarrow SL(2,\mathbb{C})\right]$}\;,
\end{split}
 \end{align}
 which follows from the fact that the quantum trace map $\textrm{tr}_q^{\mathcal{T}}$ is a module homomorphism and it becomes the classical trace map $\Phi_{\rm tr}$ in \eqref{classical trace map} at $q^{1/4}=-1$.
 
\vskip 0.3cm

\noindent
\textbf{III}.  For an even link $K$ and a {\it meridian knot}   $\bigcirc_{\rm m} $, a knot linking the  heavy knot $\mathcal{K}$,
\begin{align}
\begin{split}
\hat{O}_{K \sqcup \bigcirc^2_{\rm m}} = \hat{O}_{\bigcirc^2_{\rm m}} \cdot \hat{O}_{K} \;, 
\quad \textrm{where} \quad
\hat{O}_{\bigcirc^2_{\rm m}} := 2+ e^{\hat{\mathfrak{M}}} +   e^{-\hat{\mathfrak{M}}}  \;.\label{conjecture 6}
\end{split}
\end{align}
See \eqref{exponentiated M} for the definition of $e^{\pm \hat{\mathfrak{M}}}$ and Figure \ref{fig: Reidemeister move 1p} for the knot $\bigcirc_{\rm m}$. The  operator $\hat{O}_{\bigcirc^2_{\rm m}} $ always commutes with $e^{\hat{C}_I}$ and thus the left multiplication is well-defined in $\hat{\mathbb{C}}_q[\mathcal{T}]$. 
In section \ref{sec : perturbative invariants}, we will see that
\begin{align}
\sum_{s=0}^\infty \mathcal{Z}^{\rm (hyp)}_s (\hat{O}_{K \sqcup \bigcirc^2_{\rm m}} ) \hbar^s = 4  \sum_{s=0}^\infty \mathcal{Z}^{\rm (hyp)}_s (\hat{O}_{K} ) \hbar^s \label{Zs-under-adding-merdian-knot}
\end{align}
for an arbitrary knot $K$ in $M=S^3\backslash \mathcal{K}$. 
It is compatible  with the all-order length conjecture \eqref{conjecture 2} since
\begin{align}
\begin{split}
&J_{n, \tilde{n}_I=2}(\mathcal{K} \cup (K \sqcup \bigcirc^2_{\rm m})) =  (q^{n/2}+q^{-n/2})^2 
\times  J_{n, \tilde{n}_I=2}(\mathcal{K} \cup K ) \;,
\\
&\textrm{and } (q^{n/2}+q^{-n/2})^2|_{q=\exp (2\pi i/k), n=k} =4\;. 
\end{split}
\end{align}
Here we use the property of our Jones polynomial depicted in Figure~\ref{fig: Reidemeister move 1p} .

\section{Perturbative  knot invariants} \label{sec : perturbative invariants}

We recall and extend state-integral models from \cite{Dimofte:2011gm,Dimofte:2012qj, Gang:2015wya} 
to define the perturbative knot invariants $\{\mathcal{Z}_s^{\rm (hyp)}(\hat{O}_K;M) \}_{s=0}^\infty$.  

\subsection{State-integral model with quantum trace map} The state-integral model is based on an ideal triangulation $\mathcal{T}$  of $M= S^3\backslash \mathcal{K}$  and can be written as in the following form using Dirac brackets ($k:=|\mathcal{T}|$)
\begin{align}
\begin{split}
&Z_\hbar [M +\hat{O};X_{\bf m }] 
\\
&= \langle X_{\bf m }, C_1, \ldots , C_{k-1};\Pi_{M,C} | \hat{O}(q^{\frac{1}4}= -e^{\frac{\hbar}4}) | \Delta^{\otimes k}\rangle\big{|}_{C_1 = \ldots= C_{k-1}=0}\;,
\\
& = \int \prod_{i=1}^k \frac{dZ_i}{\sqrt{2\pi \hbar}}  \langle X_{\bf m }, C_1, \ldots , C_{k-1};\Pi_{M,C} | Z_1, \ldots, Z_k;\Pi_{Z} \rangle   \langle  Z_1, \ldots, Z_k ;\Pi_{Z} |  \hat{O} | \Delta^{\otimes k}\rangle\;.
\end{split} \label{Z[M+O]}
\end{align}
The state-integral is a function on a meridian variable $X_{\bf m}$.
Here $|X_1, \ldots, X_k;\Pi_X \rangle $ denotes the position basis of $\left( \mathcal{H}(\partial \Delta) \right)^{\otimes k}$, the Hilbert-space associated with the $k$-tetrahedra \cite{Dimofte:2011gm}, with respect to a polarization choice $\Pi_X = (X_1,\ldots, X_k , P_1, \ldots, P_k)^T$:
\begin{align}
\begin{split}
&\langle X_1, \ldots, X_k ;\Pi_X| e^{\hat{X}_i}|\psi\rangle  = e^{X_i} \langle X_1, \ldots, X_k ;\Pi_X |\psi \rangle \; \textrm{ and }
\\
&\langle X_1, \ldots, X_k ;\Pi_X | e^{\hat{P}_i}|\psi\rangle  =  \langle X_1, \ldots, X_k ;\Pi_X |\psi \rangle|_{X_i \rightarrow X_i +\hbar}, \; \forall |\psi \rangle \in \left( \mathcal{H}(\partial \Delta) \right)^{\otimes k}\;.
\end{split}
\end{align} 
Here two polarizations $\Pi_{M,C}$ and $\Pi_{Z}$ are
\begin{align}
\Pi_{M,C} = \begin{pmatrix}
\mathfrak{M}
\\
C_1
\\
\ldots
\\
C_{k-1}
\\
\frac{1}2 \mathfrak{L}
\\
\Gamma_1
\\
\ldots
\\
\Gamma_{k-1}
\end{pmatrix}\;,
\quad \Pi_{Z} = \begin{pmatrix}
Z_1
\\
Z_2
\\
\ldots
\\
Z_{k}
\\
Z''_1
\\
Z''_2
\\
\ldots
\\
Z_k''
\end{pmatrix}
\end{align}
and they are related to each other by a linear canonical transformation
\begin{align}
\Pi_{M,C} = g \cdot \Pi_{Z} - (i \pi +\frac{\hbar}2) \begin{pmatrix}\vec{\nu} \\ \vec{\nu}_p \end{pmatrix} \;, \;\; \textrm{ where } g\in Sp(2k, \mathbb{Z}) \textrm{ and } \vec{\nu},\vec{\nu}_p \in \mathbb{Z}^{k}\;. \label{g, nu in gluing eqns}
\end{align}
$\Gamma$'s are chosen such that the linear transformation becomes a canonical transformation. The choice is not unique but the final state-integral does not depend on it. 
The two  position bases are related to each other by following unitary transformation  
\begin{align}
\begin{split}
 &\langle X_{\bf m }, C_1, \ldots , C_{k-1};\Pi_{M,C} | Z_1, \ldots, Z_k;\Pi_{Z} \rangle  
 \\
 &\hskip 3cm
 = \frac{1}{\sqrt{\det B}} \exp \left( \frac{1}{2\hbar }  Q(\vec{X}, \vec{Z})\right)\bigg{|}_{\vec{X} = (X_{\mathbf{m}}, C_1,\cdots, C_{k-1})}^{\vec{Z}=(Z_1, \ldots, Z_k)}
 \\
 &\textrm{with}
 \\
 &Q(\vec{X},\vec{Z}) =  \vec{Z} B^{-1}A \vec{Z} + \vec{X} D B^{-1} \vec{X} + (2\pi i +\hbar) \vec{f}B^{-1} \vec{X} + (i \pi  + \frac{\hbar}2)^2 \vec{f} B^{-1}\vec{\nu} 
 \\
 &\qquad \qquad \;\;- \vec{Z} B^{-1} ((2i \pi +\hbar)\vec{\nu} + 2\vec{X})\;.
 \end{split}
\end{align}
Here $A,B,C$ and $D$ are the four $(k\times k)$ block matrices of $g$: 
\begin{align}
g = \begin{pmatrix}  
A & B
\\
C & D
\end{pmatrix}\;.
\end{align}
The vectors $(\vec{f},\vec{f}'') = (f_i, f_i'')_{i=1}^k $ are known as combinatorial flattening, 
and chosen to satisfy the following relation
\begin{align}
 \begin{pmatrix}  
A & B
\\
C & D
\end{pmatrix} \cdot \begin{pmatrix} \vec{f} \\ \vec{f}''\end{pmatrix} =  \begin{pmatrix}\vec{\nu} \\ \vec{\nu}_p \end{pmatrix} \;.
\end{align}
$|\Delta^{\otimes k}\rangle \in \left( \mathcal{H}(\partial \Delta ) \right)^{\otimes k}$ 
is the wave-function for $k$-tetrahedra satisfying the following difference equations
\begin{align}
(e^{\hat{Z}''_i} +e^{-\hat{Z}_i}-1)|\Delta^{\otimes k}\rangle =0 \;, \quad i =1,\ldots k.
\end{align} 
In the polarization $\Pi_{Z}$, the wave-function is  given by a product of quantum dilogarithms 
(see Appendix~\ref{App : QDL} for our convention for the quantum dilogarithm) 
\begin{align}
\langle Z_1, \ldots, Z_k ;\Pi_Z |\Delta^{\otimes k} \rangle = \prod_{i=1}^k \psi_\hbar (Z_i)\;.
\end{align}
 Gathering  all the  expressions above, one has
\begin{align}
\begin{split}
&Z_\hbar [M+\hat{O};X_{\bf m}] 
\\
&= \frac{1}{\sqrt{\det B}}  \sum_\a C_\a (q^{\frac{1}4})\big{|}_{ q^{\frac{1}4}= - e^{\frac{\hbar}4}}\int \prod_{i=1}^k \frac{dZ_i}{\sqrt{2\pi \hbar}} \exp \left(\frac{1}{2\hbar } Q(\vec{X}, \vec{Z} ) \right) \bigg{|}_{\vec{X} = (X_{\mathbf{m}}, \cdots, C_{k-1})}^{\vec{Z}=(Z_1, \ldots, Z_k)} 
\\
&\qquad \qquad \qquad  \qquad \qquad \qquad \qquad  \times \langle Z_1, \ldots, Z_k ;\Pi_Z |\prod_i \hat{z}_i^{a_i^{(\a)}}  (\hat{z}''_i)^{b_i^{(\a)}}|\Delta^{\otimes k} \rangle\;,
\\
&= \frac{1}{\sqrt{\det B}}  \sum_\a  C_\a (q^{\frac{1}4})\big{|}_{ q^{\frac{1}4}= - e^{\frac{\hbar}4}}\int \prod_{i=1}^k \frac{dZ_i}{\sqrt{2\pi \hbar}} \exp \left(\frac{1}{2\hbar } Q(\vec{X}, \vec{Z} ) \right) \bigg{|}_{\vec{X} = (X_{\mathbf{m}}, \cdots, C_{k-1})}^{\vec{Z}=(Z_1, \ldots, Z_k)} 
\\
&\qquad \qquad \qquad  \qquad \qquad \qquad \qquad  \times\prod_{i=1}^k \psi_\hbar (Z_i +b_i^{(\a)}\hbar) \exp \left( a_i^{(\a)} Z_i\right)\;,
\end{split}
\end{align}
when $\hat{O} = \sum_{\a} C_\a(q^{\frac{1}4})\prod_i \hat{z}_i^{a^{(\a)}_i} (\hat{z}''_i)^{b^{(\a)}_i}  \in \hat{\mathbb{C}}_q[\mathcal{T}]$.  From the expression in \eqref{Z[M+O]}, it is not difficult to see that the $Z_{\hbar}[M+\hat{O}; X_{\bf m }]$ with $\hat{O} \in \hat{\mathbb{C}}_q[\mathcal{T}]$ is well-defined (recall the definition of $\hat{\mathbb{C}}_q[\mathcal{T}]$ in \eqref{quantum gluing module}), i.e.
\begin{align}
Z_\hbar [M +\hat{O}] = Z_\hbar [M + e^{\hat{C_I}}\cdot \hat{O}] = Z_\hbar [M+\hat{O}\cdot (\hat{z}_i^{-1}+\hat{z}''_i) ]
\end{align}
It is also straightforward to see that
\begin{align}
Z_\hbar [M+F(e^{\hat{\mathfrak{M}}}) \cdot \hat{O}] = F(e^{X_{\bf m}}) Z_{\hbar }[M+\hat{O}]\;, \label{Z[M+F(eM)]}
\end{align}
for arbitrary Laurent polynomial $F(z)$ and $\hat{O} \in \hat{\mathbb{C}}_q[\mathcal{T}]$.

\subsection{Perturbative invariants } 
We are ready to define and explain the perturbative invariants $\{\mathcal{Z}^{(\rm hyp)}_s(\hat{O};M)\}_{s=0}^{\infty}$.
By expanding the state-integral in the limit $\hbar \rightarrow 0$ around the saddle point $\vec{Z}^{(\rm hyp)}$ associated to complete hyperbolic structure of $M$, one have  
\begin{align}
Z_\hbar [M+\hat{O};X_{\bf m}=0]  \xrightarrow{ \quad \hbar \rightarrow 0 \quad } \exp \left( \sum_{n=0}^{\infty} S_n^{(\rm hyp)} (\hat{O};M) \hbar^{n-1}\right)\;. \label{perturbative Sna}
\end{align}
In the expansion, one can use the asymptotic expansion of the quantum dilogarithm function in \eqref{eq:psi expansoin}. 
Then, we define the perturbative invariants $\{\mathcal{Z}^{(\rm hyp)}_s,\mathcal{W}^{(\rm hyp)}_s \}_{s=0}^\infty$ as
\begin{align}
\begin{split}
    \frac{\exp \left( \sum_{n=0}^{\infty} S_n^{(\rm hyp)} (\hat{O};M) \hbar^{n-1}\right) }{\exp \left( \sum_{n=0}^{\infty} S_n^{(\rm hyp)} (M) \hbar^{n-1}\right)} &= \sum_{s=0}^{\infty} \mathcal{Z}_s^{(\rm hyp)} (\hat{O};M) \hbar^{s}
\\
&= \exp \left(   \sum_{s=0}^{\infty} \mathcal{W}_s^{(\rm hyp)} (\hat{O};M) \hbar^{s} \right)  \label{perturbative Zs}\;.
\end{split}
\end{align}
Here $S_n^{\rm (hyp)}(M):=S_n^{\rm (hyp)} (\hat{O}_{K=\emptyset};M)$ is the perturbative invariant of state-integral model without any insertion of loop operator.   From the definition, one can see  that the relation in \eqref{Zs-under-adding-merdian-knot} simply follows from  \eqref{Z[M+F(eM)]} with $F(x)=2+x+\frac{1}x$ and $X_{\bf m}=0$.  In the classical limit $\hbar \rightarrow 0$,  saddle points $\{\vec{Z}=\vec{Z}^{(\alpha)}\}$ of the state-integral at $X_{\bf m}=0$ satsify following equations
\begin{align}
\prod_{j=1}^k z_j^{A_{ij}} (1-z^{-1}_{j})^{B_{ij}}\big{|}_{z_i= \exp( Z_i^{(\alpha)} )}= (-1)^{\nu_i}\;.
\end{align}
For $\rho_{\rm hyp}$-regular ideal triangulation $\mathcal{T}$, the saddle point $\vec{Z}^{(\rm hyp)}$ is uniquely characterized by  following conditions ($z^{(\rm hyp)}_i : =\exp (Z^{(\rm hyp)}_i)$)
\begin{align}
\textrm{Im}[z^{(\rm hyp)}_i]<0 \quad \textrm{for all }i=1,\ldots, k\;. \label{Zhyp}
\end{align}

\example[]{As an example, consider the case with $M = S^3\backslash \mathbf{4}_1$.  Its simplest ideal triangulation consists of two tetrahedra, say $\Delta_Y$ and $\Delta_Z$, with edge parameters 
\[
(y,y',y'')=(e^Y, e^{Y'},e^{Y''}) \;\; \mbox{and} \;\;  (z,z',z'')=(e^{Z},e^{Z'},e^{Z''}). 
\]
Using the gluing data given in   Appendix \ref{gluing}, we have
\begin{align}
\begin{pmatrix}
\mathfrak{M}
\\
C_1
\\
\frac{1}2 \mathfrak{L}
\\
\Gamma_1
\end{pmatrix} = \begin{pmatrix}
0 & -1 & 1 & 0 
\\
1 & 1 & -1 & -1
\\
0 & 1 & 0 & -1 
\\
0 & 0 & 1 & 0 
\end{pmatrix} \cdot \begin{pmatrix}
Y
\\
Z
\\
Y''
\\
Z''
\end{pmatrix}\,.
\end{align}%
The corresponding state-integral model is \cite{Dimofte:2011gm,Dimofte:2012qj} 
\begin{align}
\begin{split}
&Z_{\hbar} (S^3\backslash \mathbf{4}_1 ; X_{\bf m}) = \langle X_{\bf m} , C_1 = 0 | \Delta^{\otimes 2} \rangle\;,
\\
& \qquad =  \int \frac{dY dZ}{2\pi \hbar} \exp \left(\frac{1}{2\hbar} (X_{\bf m}^2 - 2 X_{\bf m} Y+2X_{\bf m} Z - 2 Y Z ) \right) \psi_{\hbar}(Y)\psi_{\hbar}(Z)\;.
\end{split}
\end{align}
The saddle points of the integral are the solutions of the following equations:
\begin{align}
\begin{split}
&\partial_{Y,Z} \left(\textrm{Li}_2 (e^{-Y})+\textrm{Li}_2 (e^{-Z}) + \frac{1}2 X_{\bf m}^2 +X_{\bf m} Z- X_{\bf m} Y - Y Z\right) =0 \;,
\\
&\Longrightarrow \quad -Z+\log (1-e^{-Y}) = X_{\rm m}\;, \quad -Y+\log (1-e^{-Z}) = -X_{\mathbf{m}}\;. \label{saddle points}
\end{split}
\end{align}
The saddle point equations coincide with the (logarithmic) gluing equations of the ideal triangulation. 
At $X_{\bf m} =0$, there are two saddle points,
\begin{align}
\begin{split}
&(\alpha) = (\overline{ \textrm{hyp}}) \;:\;  Y = Z = \frac{i \pi}3 \;,
\\
&(\alpha) = ( \textrm{hyp}) \;:\;  Y = Z =- \frac{i \pi}3 \;. \label{41-saddle-points}
\end{split}
\end{align}

We now extend the state-integral model for $S^3\backslash \mathbf{4}_1$ to include a quantum trace map $\hat{O}_{K_b}$, where $K_b$ is the geodesic knot in the homotopy class $ b \in \pi_1 (S^3\backslash \mathbf{4}_1)$, see Figure \ref{fig: figure8-fundamental}. 
The classical trace map is
\begin{align}
\Phi_{\rm tr} (Y_{K_b}) = -(y^{-1} +z^{-1}-y^{-1} z^{-1}) \in \mathbb{C}[\chi[\mathcal{T}]]\;.
\end{align}
The expression is obatined using the holonomy matrices in \eqref{hol matrix} and the  gluing equation varieity $\chi[\mathcal{T}]$ is given in \eqref{gluing equation variety of 41}.
After quantization, we assume that the quantum loop operator  is given as 
\begin{align}
\hat{O}_{K_b} = \textrm{tr}_q (Y_{K_b})=-C_1 \hat{y}^{-1} -C_2 \hat{z}^{-1}  +C_3 \hat{y}^{-1} \hat{z}^{-1}  \in \hat{\mathbb{C}}_q[\mathcal{T}]
\end{align}
where $\{C_{\a=1,2,3} \}$ are Laurent polynomials in $q^{1/4}$ which all become $1$ in the limit $q^{1/4}\rightarrow -1$. 
Later in section \ref{sec : figure-eight knot}, we will propose that  $C_1 = C_2 =C_3= q^{1/2} = e^{\frac{\hbar }2}$ and check  it  against the all-order length conjecture. 
Under the quantization,  the state-integral is 
\begin{align}
\begin{split}
&Z_{\hbar} \left[ S^3\backslash \mathbf{4}_1+\hat{O}_{K_b} ; X_{\bf m}=0\right] 
\\
&\; =  \int \frac{dY dZ}{2\pi \hbar}  \exp \left(-\frac{YZ}{\hbar}  \right)  \psi_{\hbar}(Y)\psi_{\hbar}(Z) \times e^{\hbar/2}\left( -e^{-Y} -e^{-Z}+e^{-Y-Z} \right)\;. \label{SI-41-Kb}
\end{split}
\end{align}
For $K_b^{ 2}$ (2-cabling of $K_b$), the classical trace map is 
\begin{align}
\Phi_{\rm tr} (Y_{K_b^2}) =\left( \Phi_{\rm tr} (Y_{K_b}) \right)^2  = (y^{-1}+z^{-1}-y^{-1} z^{-1})^2 \in \mathbb{C}[\chi_0[\mathcal{T}]]\;.
\end{align}
After quantization, as will be proposed in section \ref{sec : figure-eight knot}, the quantum loop operator is
\begin{align}
\hat{O}_{K_b^{ 2}} = q (\hat{y}^{-1}  + \hat{z}^{-1}  - \hat{y}^{-1} \hat{z}^{-1} )^2 +1-q^2  \in \hat{\mathbb{C}}_q[\mathcal{T} ]\;.
\end{align}
The state-integral with the loop operator is 
\begin{align}
\begin{split}
&Z_{\hbar} \left[ S^3\backslash \mathbf{4}_1+\hat{O}_{K_b^{ 2}} ; X_{\bf m}=0\right] 
\\
&\; =  \int \frac{dY dZ}{2\pi \hbar}  \exp \left(-\frac{YZ}{\hbar}  \right)  \psi_{\hbar}(Y)\psi_{\hbar}(Z) \times \bigg{(}  e^{\hbar}\left(e^{-Y}+e^{-Z}-e^{-Y-Z}\right)^2+1-e^{2\hbar} \bigg{)}\;.  \label{SI-41-Kb2}
\end{split}
\end{align}}
\subsection{Feynman perturbation theory for the invariants}

In this section we will give an expression for the invariants $\{\mathcal{W}_s^{\rm hyp}(\hat{O}_K;M) \}_{s=0}^\infty$ using Feynman diagrams following \cite{Dimofte:2012qj}.The quantum loop operator $\hat{O}$ can be generally given in the following form

\begin{align}
	\hat{O}= \sum_{\a} C_\alpha (q^{1/4})\hat{O}_{\a} = \sum_\a C_\a (q^{1/4}) \prod_i \hat{z}_i^{a^{(\a)}_i} (\hat{z}''_i)^{b^{(\a)}_i}\;.
\end{align} 
Giving an explicit formula for each of the terms $\mathcal{W}_s^{\rm hyp}(\hat{O}_K;M)$ is quite involved. We will instead give an implicit expression in terms of the perturbative expansion of each term $Z_{\hbar}(M+\hat{O}_{\alpha};X_{\mathbf{m}})$ in the expression
\begin{eqnarray}
Z_{\hbar}(M+	\hat{O};X_{\mathbf{m}})=\sum_{\alpha}C_{\alpha}(q^{\frac{1}4} )Z_{\hbar}(M+\hat{O}_{\alpha};X_{\mathbf{m}})\big{|}_{q^{\frac{1}4} = -e^{\frac{\hbar}4}}\,,
\end{eqnarray}
which we denote them as $S^{(c)}_{n;\alpha}$:
\begin{align}
& Z_{\hbar}(M+\hat{O}_{\alpha};X_{\mathbf{m}})
\nonumber \\
&=\frac{1}{\sqrt{\mathrm{det}B}}\int \prod_{i=1}^{k}\frac{dZ_{i}}{\sqrt{2\pi \hbar}}\exp\left(\frac{1}{2\hbar}Q(Z,X_{\mathbf{m}})\right)e^{a^{(\alpha)}\cdot Z}\prod_{i=1}^{k}\psi_{\hbar}(Z_{i}+b^{(\alpha)}_{i}\hbar)
\nonumber \\
&\xrightarrow{\;\textrm{around a saddle point $Z^{(c)}$ in $\hbar \rightarrow 0$} \;} \exp\left(\sum_{n=0}^{\infty}\hbar^{n-1}S^{(c)}_{n;\alpha}\right).
\end{align}
The expansion can be obtained by considering the expansion $Z_{i}=Z^{(c)}_{i}+Y_{i}$.  So, the integral expression becomes
\begin{align}
Z_{\hbar}(M+\hat{O}_{\alpha};X_{\mathbf{m}})=\frac{e^{\Gamma^{(0)}}}{\sqrt{\mathrm{det}B}}\int \prod_{i=1}^{k}\frac{dY_{i}}{\sqrt{2\pi \hbar}}e^{\frac{1}{2\hbar}Y_{i}H^{ij}Y_{j}}\prod_{i=1}^{k}e^{\sum_{s=1}^{\infty}\Gamma^{(s)}_{i}Y_{i}^{s}/s!},
\end{align}
where we defined
\begin{eqnarray}
H^{ij}:=(B^{-1}A)_{ij}+(z_{j}z_{j}'')^{-1}\delta_{ij},
\end{eqnarray}
the `vacuum energy':
\begin{eqnarray}
\Gamma^{(0)}=F(X_{\mathbf{m}},Z^{(c)})+\sum_{i=1}^{k}\sum_{n=1}^{\infty}\frac{B_{n}(1+b_{i}^{(\alpha)})\hbar^{n-1}}{n!}\mathrm{Li}_{2-n}\left(e^{-Z_{i}^{(c)}}\right),
\end{eqnarray}
where $B_{n}(1+b_{i}^{(\alpha)})$ denotes the Bernoulli polynomials evaluated at $1+b_{i}^{(\alpha)}$ and
\begin{eqnarray}
F(X_{\mathbf{m}},Z^{(c)}):=\frac{1}{2\hbar}Q(Z^{(c)},X_{A})+a\cdot Z^{(c)}+\frac{1}{\hbar}\sum_{i}\mathrm{Li}_{2}(e^{-Z_{i}^{(c)}}),\nonumber
\end{eqnarray}
the linear vertex
\begin{eqnarray}
\Gamma^{(1)}_{i}=a^{(\alpha)}_{i}+\frac{1}{2}-\sum_{n=1}^{\infty}\frac{B_{n}(1+b_{i}^{(\alpha)})\hbar^{n-1}}{n!}\mathrm{Li}_{1-n}\left(e^{-Z_{i}^{(c)}}\right),
\end{eqnarray}
and higher valence vertices ($k\geq 2$):
\begin{align}
\Gamma^{(k)}_{i}=(-1)^{k}\sum_{n=0}^{\infty}\frac{B_{n}(1+b_{i}^{(\alpha)})\hbar^{n-1}}{n!}\mathrm{Li}_{2-n-k}\left(e^{-Z_{i}^{(c)}}\right).
\end{align}
Then, the coefficients $S^{(c)}_{n;\alpha}$ can be computed as sums over Feynman diagrams corresponding to non-directed graphs, without open edges, which we denote as $\mathcal{G}_{\Gamma}$. The situation is completely analogous to \cite{Dimofte:2012qj}, with a slight modification on the valence of the vertices. We associate a weight to each $\mathcal{G}_{\Gamma}$:
\begin{eqnarray}
W(\mathcal{G}_{\Gamma}):=\frac{1}{|\mathrm{Aut}(\mathcal{G}_{\Gamma})|}\sum_{\mathrm{labels}}\prod_{v\in \mathrm{vertices}}\Gamma^{(k_{v})}_{l_{v}}\prod_{e\in\mathrm{edges}}\Pi_{e}
\end{eqnarray}
here $l_{v}$ is the set of labels corresponding to the legs of the vertex $v$ in the graph and $k_{v}$ its valence. Given a graph $\mathcal{G}_{\Gamma}$ we denote $\mathcal{L}$ the number of loops, $V_{a}$ the number of vertices of valence $a$. We define 
\begin{eqnarray}
\mathcal{G}_{n}:=\{\mathcal{G}_{\Gamma}:\mathcal{L}+V_{1}+V_{2}\leq n\}
\end{eqnarray}
then, the $S^{(c)}_{n;\alpha}$ coefficients are given by
\begin{eqnarray}
S^{(c)}_{0;\alpha}=\mathrm{coeff}[\Gamma^{(0)},\hbar^{-1}],
\end{eqnarray}
the 1-loop term:
\begin{eqnarray}
e^{S^{(c)}_{1,\alpha}}=\frac{e^{\mathrm{coeff}[\Gamma^{(0)},\hbar^{0}]}}{\sqrt{\mathrm{det}B}}\frac{i^{k}}{\sqrt{\mathrm{det}H}},
\end{eqnarray}
and for $n\geq 2$:
\begin{eqnarray}
S^{\alpha}_{n,k}=\mathrm{coeff}\left\{\Gamma^{(0)}+\sum_{\mathcal{G}_{\Gamma}\in\mathcal{G}_{n}}W(\mathcal{G}_{\Gamma}),\hbar^{n-1}\right\}\qquad n\geq 2.
\end{eqnarray}
Now we can use these results to compute the invariants $\{\mathcal{W}_s^{\rm hyp}(\hat{O};M) \}_{s=0}^\infty$. First denote
\begin{eqnarray}\label{expansionformal}
&&\frac{Z_{\hbar}(M+\hat{O};X_{\mathbf{m}})}{Z_{\hbar}(M;X_{\mathbf{m}})} = \frac{1}{Z_{\hbar}(M;X_{\mathbf{m}})}\sum_{\alpha}C_{\alpha}(q^{\frac{1}4})Z_{\hbar}(M+\hat{O}_{\alpha};X_{\mathbf{m}})\nonumber\\
&&\xrightarrow{\;\textrm{around a saddle point $Z^{(c)}$ in $\hbar \rightarrow 0$} \;} \sum_{\alpha}C_{\alpha}
(q^{\frac{1}4}=-e^{\frac{\hbar}4})\exp\left(\sum_{n=1}^{\infty}\hbar^{n-1}\widetilde{S}^{(c)}_{n,\alpha}
\right)
\end{eqnarray}
where
\begin{eqnarray}
\widetilde{S}^{(c)}_{n,\alpha}:=S^{(c)}_{n,\alpha}-S^{(c)}_{n}(M)\qquad n\geq 1
\end{eqnarray}
and we have used the fact that
\begin{eqnarray}
S^{(c)}_{0}(M)=S^{(c)}_{0,\alpha}\qquad \text{for all \ \ }\alpha.
\end{eqnarray}
Therefore, the expression of $\widetilde{S}^{(c)}_{n,\alpha}$, $n\geq 1$ in terms of Feynman diagrams reduce exactly to the same computation than $S^{(c)}_{n}(M)$, presented in \cite{Dimofte:2012qj} with two modifications in the vertices:
\begin{enumerate}
\item The vertices of valence $0$ and $1$ becomes
\begin{eqnarray}
\Gamma^{(0)} &=& a^{(\alpha)}\cdot Z^{(c)}+\sum_{i=1}^{k}\sum_{n=1}^{\infty}\frac{(B_{n}(1+b_{i}^{(\alpha)})-B_{n})\hbar^{n-1}}{n!}\mathrm{Li}_{2-n}\left(e^{-Z_{i}^{(c)}}\right),\nonumber\\
\Gamma^{(1)}_{i}&=& a^{(\alpha)}_{i}-\sum_{n=1}^{\infty}\frac{(B_{n}(1+b_{i}^{(\alpha)})-B_{n})\hbar^{n-1}}{n!}\mathrm{Li}_{1-n}\left(e^{-Z_{i}^{(c)}}\right),
\end{eqnarray}
 \item The coefficient $B_{n}=B_{n}(1)$ is replaced by $B_{n}(1-b^{(\alpha)}_{i})-B_{n}$ in the vertex $\Gamma_{i}^{(s)}$, $s\geq 2$
\end{enumerate}

The coefficients $C_{\alpha}$ are Laurent polynomials in $q^{1/4}$, so they can be expanded in power series in $\hbar$:
\begin{eqnarray}
C_{\alpha}(q^{\frac{1}4} =- e^{\frac{\hbar }4})=\sum_{s\geq 0}\hbar^{s}f_{s,\alpha}\,.
\end{eqnarray}
Therefore, the series expression in (\ref{expansionformal}) can be written as a formal power series in $\hbar$ whose $\mathcal{O}(\hbar^{0})$ term is given by
\begin{eqnarray}\label{logexp}
\frac{Z_{\hbar}(M+\hat{O};X_{\mathbf{m}})}{Z_{\hbar}(M;X_{\mathbf{m}})}=\sum_{c}W^{(c)}_{\gamma}+\mathcal{O}(\hbar)
\end{eqnarray}
where $W^{(c)}_{K}:=\sum_{\alpha}f_{0,\alpha}e^{\widetilde{S}^{(c)}_{1,\alpha}}$ is the classical expectation value of the light knot $K$ at the saddle point $(c)$. Therefore, for a fixed saddle point $(c)$  we can write (\ref{logexp}) as
\begin{eqnarray}
\ln\left( W^{(c)}_{K}\right)+\ln\left(1+P^{(c)}(\hbar)\right)\,,
\end{eqnarray}
where $P^{(c)}(\hbar)$ is a power series in $\hbar$, without a constant term. Hence, the Taylor series of the logarithm defines the invariants $\{\mathcal{W}_s^{\rm (hyp)}(\hat{O}_K;M) \}_{s=0}^\infty$. They become very involved as $s$ grows larger and we find it not very illuminating to include a general formula here. Instead we we will present explicit expressions for $s=0,1$. 

For $s=0$, we get the classical expectation value, as remarked above
\begin{align}
\begin{split}
   W_{K}^{\mathrm{class}}:=\exp\left(\mathcal{W}_0^{(\mathrm{hyp})}(\hat{O};M)\right)
&=\sum_{\alpha}f_{0,\alpha}e^{\widetilde{S}^{(\mathrm{hyp})}_{1,\alpha}}
\\
&=\sum_{\alpha}C_{\alpha}(q^{1/4}=-1)\prod_{i}(z''_{i})^{b^{(\alpha)}_{i}}z_{i}^{a^{(\alpha)}_{i}} \,, 
\end{split}
\end{align}
where $z''_{i}=(1-e^{-Z_i})$ and $z_{i}=e^{Z_i}$ are evaluated at the saddle point $Z^{(\mathrm{hyp})}$ in \eqref{Zhyp}. Hence the value of $\exp\left(\mathcal{W}_0^{(\mathrm{hyp})}(\hat{O};M)\right)$ coincides with the classical expectation value of $\hat{O}$, as expected. 

For $s=1$, we get
\begin{eqnarray}
\mathcal{W}_1^{\rm (hyp)}(\hat{O};M)=\frac{\sum_{\alpha}(f_{1,\alpha}+f_{0,\alpha}
\widetilde{S}^{\mathrm{(hyp)}}
_ { 2,\alpha } )e^{\widetilde{S}^{\mathrm{(hyp)}}_{1,\alpha}} }{W_{K}^{\mathrm{class}}} \,,
\end{eqnarray}
where the explicit expressions for $\widetilde{S}^{\mathrm{(hyp)}}
_ { 1,\alpha }$, $\widetilde{S}^{\mathrm{(hyp)}}
_ { 2,\alpha }$ can be obtained with the help of the Feynman diagram expansion as described above. In the next section we will provide explicit exampled for links in the figure-eight knot-complement.

\section{Example : figure-eight knot-complement } \label{sec : figure-eight knot}
We explicitly define the quantum trace map for $M=S^3\backslash \mathbf{4}_1$ and prove (or numerically check) some parts of conjecture in \eqref{main conjecture}.    

\begin{figure}[htbp]
	\begin{center}
		\includegraphics[width=.3\textwidth]{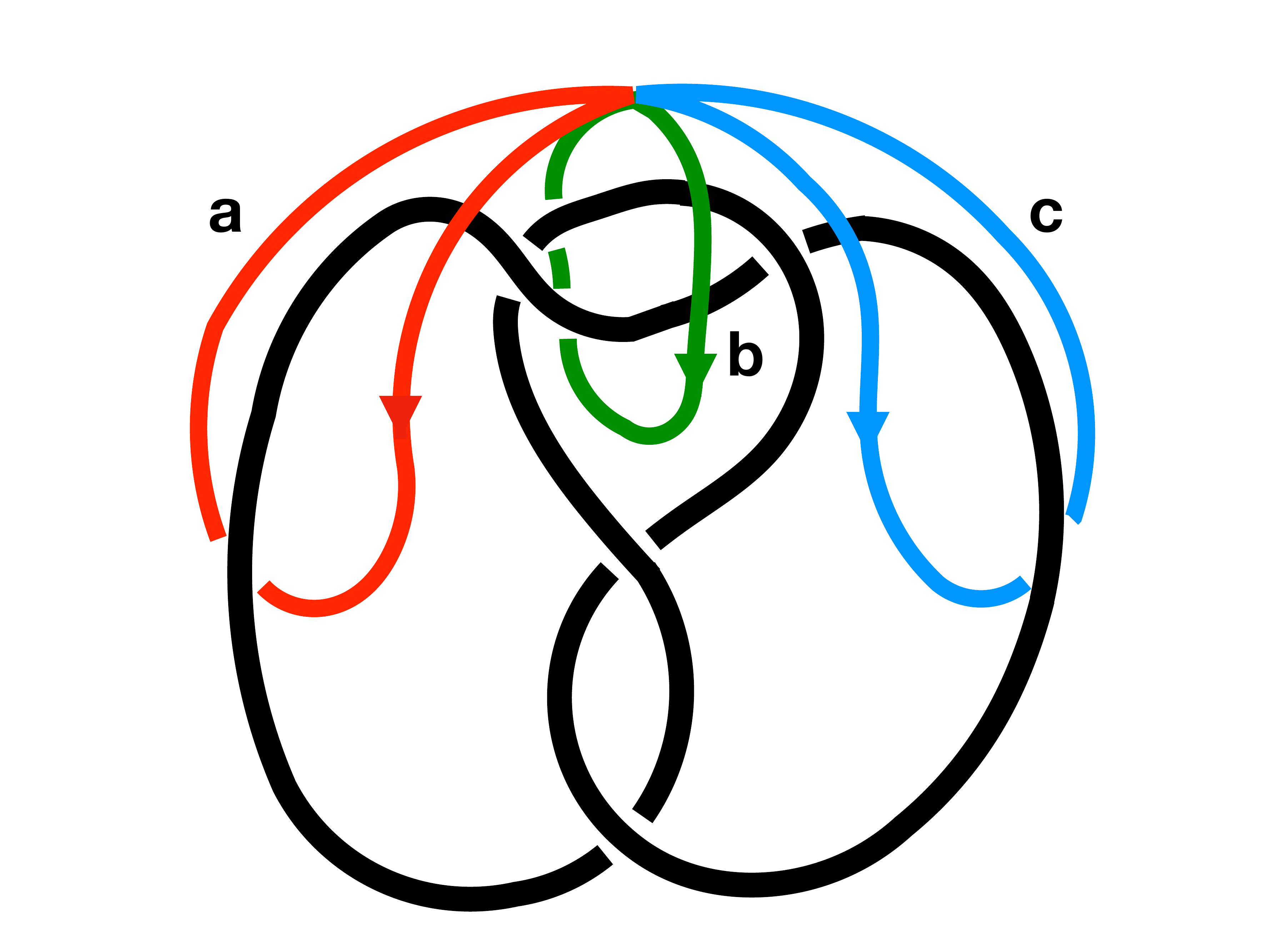}
	\end{center}
	\caption{Three generators of fundamental group, given in \eqref{pi1-presentation}, of the figure-eight knot complement. }
	\label{fig: figure8-fundamental}
\end{figure}

The Skein module for $M=S\backslash \mathbf{4}_1$ is studied in \cite{bullock2005kauffman}:
\begin{align}
\begin{split}
&\mathcal{S}_{q}(S^3\backslash \mathbf{4}_1) = \mathbb{C}[q^{\pm 1/4}]\textrm{-module with basis $\{Y_{K_b^{ n}\sqcup \bigcirc_{\rm m}^m}\}_{m \geq 0, 0\leq n \leq 2 }$} 
\\
&\Rightarrow  
\\
&\mathcal{S}^{\rm even}_{q}(S^3\backslash \mathbf{4}_1)   = \mathbb{C}[q^{\pm 1/4}]\textrm{-module with basis  $\{Y_{K_b^{ n}\sqcup \bigcirc_{\rm m}^{2m}}\}_{m \geq 0, 0\leq n \leq 2 }$} 
\end{split}
\end{align}
Where $\bigcirc_{\rm m}$ is the meridian knot, i.e. the knot in the red color,  and $K_b$ is the knot in green color depicted in Figure~\ref{fig: figure8-fundamental}.   $K_b$ is an even knot while $\bigcirc_{\rm m}$ is an odd knot. 
$K^n$ denotes the $n$-cabling of a knot $K$. 

\subsection{Explicit  quantum trace map}
The quantum gluing module $\hat{\mathbb{C}}_q [\mathcal{T}]$ for an ideal triangulation $\mathcal{T}$ in Appendix \ref{gluing} is
\begin{align}
\begin{split}
&\hat{\mathbb{C}}_q [\mathcal{T} ] =\frac{\mathbb{C}[q^{\pm 1/4}][\hat{y}^{\pm 1}, \hat{z}^{\pm 1}, (\hat{y}'')^{\pm 1}, (\hat{z}'')^{\pm 1}]\big{/}\langle \hat{z} \hat{z}''- q^{-1} \hat{z}'' \hat{z}, \hat{y}\hat{y}'' - q^{-1} \hat{y}'' \hat{y}\rangle}{\textrm{equivalence relations}\sim} \;, 
\\
&\textrm{where }\hat{O} \sim q^{-1} \hat{y} \hat{z} (\hat{y}''\hat{z}'')^{-1} \hat{O}\;, \; \hat{O} \sim \hat{O}\cdot (\hat{y}^{-1} + \hat{y}'')\; \textrm{and}\;\hat{O} \sim \hat{O}\cdot (\hat{z}^{-1} + \hat{z}'') \;.
\end{split}
\end{align}
We propose following  quantum trace map $\textrm{tr}_q^{\mathcal{T} } \;:\;\mathcal{S}^{\rm even}_q [M] \rightarrow \hat{\mathbb{C}}_q [\mathcal{T}]$
\begin{align}
\begin{split}
&\textrm{tr}^{\mathcal{T}}_q (Y_{\bigcirc_{\rm m}^{2m}}) = \left(\hat{y}''/\hat{z} + \hat{z}/\hat{y}'' +2\right)^{m}   \;,
\\
&\textrm{tr}^{\mathcal{T}}_q (Y_{ K_b \sqcup \bigcirc_{\rm m}^{2m}}) =   -q^{1/2}\left(\hat{y}''/\hat{z} + \hat{z}/\hat{y}'' +2\right)^{m}   \left( \hat{y}^{-1} + \hat{z}^{-1} -\hat{y}^{-1}\hat{z}^{-1}\right)\;,
\\
&\textrm{tr}^{\mathcal{T}}_q (Y_{ K_b^2 \sqcup \bigcirc_{\rm m}^{2m}})= \left(\hat{y}''/\hat{z} + \hat{z}/\hat{y}'' +2\right)^{m} \left(  q(  \hat{y}^{-1} + \hat{z}^{-1} -\hat{y}^{-1}\hat{z}^{-1} )^2-q^2+1 \right)\;. \label{trace map for 41}
\end{split}
\end{align}
Assuming  the conjecture III  in  \eqref{main conjecture} and using the fact $e^{\hat{\mathfrak{M}}} = \hat{y}''/\hat{z}$, the  only non-trivial parts of the proposal are $\textrm{tr}^{\mathcal{T}}_q (Y_{K_b})$ and $\textrm{tr}^{\mathcal{T}}_q (Y_{K_b^2})$. We support the proposal by checking the all-order length conjecture.  

The perturbative invariant $\mathcal{W}^{\rm (hyp)}_s (\hat{O}_K ; S^3\backslash \mathbf{4}_1)$  defined in \eqref{perturbative Zs} for $K=K_b$ and $K=K_b^2$ can be computed by expanding the state-integral models in \eqref{SI-41-Kb} and \eqref{SI-41-Kb2} around the saddle point in \eqref{41-saddle-points}. Up to $s=2$, one has
\begin{align}
	\begin{split}
		&\sum_{s=0}^{2} \mathcal{W}^{\rm (hyp)}_s (\hat{O}_{K_b};S^3\backslash \mathbf{4}_1) \hbar^{s} = \frac{7i \pi + 3\log 3}{6} + \frac{i \sqrt{3} + 6}{18} \hbar + \frac{i}{18 \sqrt{3}} \hbar^2 \;,
		\\
		&\sum_{s=0}^{2} \mathcal{W}^{\rm (hyp)}_s (\hat{O}_{K_b^{ 2}};S^3\backslash \mathbf{4}_1) \hbar^{s} = \frac{i \pi + 3\log 3}{3}  + \frac{i\sqrt{3}+2}3 \hbar + \frac{i \sqrt{3}+6}{27} \hbar^2\;.
	\end{split}
	\label{2-loop-summary}
\end{align}
The leading part is related to the complex length of  $\gamma_b$ as follows
\begin{align}
	\begin{split}
		\mathcal{W}^{\rm (hyp)}_0 (\hat{O}_{K_b};S^3\backslash \mathbf{4}_1) &=\log \left( -\left(e^{\ell_\mathbb{C}(\gamma_b)/2} +e^{-\ell_\mathbb{C}(\gamma_b)/2}\right) \right)
		\\
		&= \frac{7i \pi + 3 \log 3}{6} \simeq 0.549306\, +3.66519 i \;.
		\\
		\mathcal{W}^{\rm (hyp)}_0 (\hat{O}_{K_b^{ 2}};S^3\backslash \mathbf{4}_1) &= 2 \mathcal{W}_0 (\hat{O}_{K_b};S^3\backslash \mathbf{4}_1) \; (\textrm{mod } 2\pi i )
		\\
		&\simeq 1.09861\, +1.0472 i\;. \label{classical W0s}
	\end{split}
\end{align}

\subsection{Test of the length conjecture - analytic proof up to 1-loop}

For the figure-8 knot complement, one of the proofs of the classical volume conjecture \cite{murakami2011introduction} is based on the Habiro-L\^e formula \cite{habiro2002quantum, masbaum2003skein} for the Jones polynomial:
\begin{align}
	\begin{split}
		J_n(q) &:= \frac{J_n (K=\mathbf{4}_1;q)}{J_n (\bigcirc;q)}
		\\
		&=\sum_{k=0}^{n-1} \prod_{j=1}^k \left(q^{(n-j)/2} - q^{-(n-j)/2} \right)  
		\left(q^{(n+j)/2} - q^{-(n+j)/2} \right) = \sum_{k=0}^{n-1} w_{n,k}(q) \,.
		\label{HL2a}
	\end{split}
\end{align}
At $q=e^{2\pi i/n}$, the function $w_{n,k}(q)$ simplifies quite a bit: 
\begin{align}
	w_{n,k} (q = e^{2\pi i/n}) = \prod_{j=1}^k 4\sin^2\left(\frac{j\pi}{n}\right) \,.
\end{align}
In the remainder of this section, $w_{n,k}$ will mean this this product depending on $n$ and $k$. 
Since $w_{n,k}$ is positive definite, it makes sense to regard $J_n^*  := J_n(q = e^{2\pi i/n})$ as a sort of partition function, 
and use it to define a probability distribution:
\begin{align}
	p_{n,k} := \frac{w_{n,k}}{J_n^*} \,, 
	\quad 
	p_{n,k} \ge 0 \,,
	\quad 
	\sum_{k=0}^n p_{n,k} = 1 \,.
	\label{prob-int}
\end{align}
(The two boundary cases, $k=0$ and $k=n$, have zero probability.)

We provide an analytic proof of the length conjecture for the figure 8-knot complement 
in the presence of the $K_b$ or $K_b^2$. 
In this context, the length conjecture is a prediction on the large $n$ limit of the series, 
\begin{align}
	A_{n}^{K_b} = \log \left( \frac{J_{n,2}^*}{J_n^*} \right) \,,
	\quad 
	A_{n}^{K_b^2} = \log \left( 1+ \frac{J_{n,3}^*}{J_n^*}\right) \,.
\end{align}
Our proof relies on simple formulas for $J_{n,2}(q)$ and $J_{n,3}(q)$: 
\begin{align}
\begin{split}
    J_{n,2}(q) &:= \frac{(-1)\times J_{n,\tilde{n}=2} (K=\mathbf{4}_1 \cup K_b;q)}{J_n (\bigcirc;q)} 
	\\
	&=  q^{1/2} J_n(q) + q^{-1/2} \sum_{k=0}^{n-1} q^{-k} w_{n,k}(q) \,, 
	\label{Jn2}
\end{split}
	\\
\begin{split}
	J_{n,3}(q) &:= \frac{ J_{n,\tilde{n}=3} (K=\mathbf{4}_1 \cup K_b;q)}{J_n (\bigcirc;q)}   
	\\
	&=(1+q) \sum_{k=0}^{n-1} q^{-k} w_{n,k}(q) + q^{-1} \sum_{k=0}^{n-1} q^{-2k} w_{n,k}(q) \,.
	\end{split}
	\label{Jn3}
\end{align}
For $J_{n,2}$, we include a sign factor to compensate the sign factor appearing in \eqref{sign factor}. 
After the compensation,  it becomes identical to the conventional Jones polynomial. These formulas are {\it conjectural}. 
We have verified that they agree with computations based on braid diagrams and $R$-matrices up to $n=20$. 
It would be interesting to derive these formulas from first principles, 
for example, using recursion relations \cite{garoufalidis2005colored} 
which proved extremely useful for Jones polynomials for single-component knots. 

With the probability interpretation of \eqref{prob-int}, the series $A_n$ are rephrased as 
\begin{align}
	\begin{split}
		A_{n}^{K_b} &= \log\left( q^{1/2} + q^{-1/2} E_{n,1} \right)  \,, 
		\quad 
		E_{n,a} = \sum_{k} q^{-ak} p_{n,k} \,, 
		\\
		A_{n}^{K_b^2} &= \log \left( 1+ (1+q) E_{n,1} + q^{-1} E_{n,2} \right)\,.
	\end{split}
\end{align}
We recognize $E_{n,a}$ as the expectation value of the function $q^{-ak}$ with respect to the 
probability distribution $p_{n,k}$. 

As explained in {\it e.g.} \cite{murakami2011introduction}, $w_{n,k}$ for fixed $n$ reaches its maximum value at $k/n = 5/6$.  
As $n$ increases, the ratio of the maximum value to generic values becomes exponentially large. 
So, as in a typical problem in statistical mechanics, we can approximate the probability distribution 
as Gaussian. 

At the classical $(s=0)$ level of the conjecture, 
the only information we need is the position of the peak of the Gaussian:  
\begin{align}
	E_{n,1} \approx q^{-k} |_{(k/n=5/6)}  = e^{-5\pi i/3} \,, 
	\quad 
	E_{n,2} \approx q^{-2k} |_{(k/n=5/6)}  = e^{-4\pi i/3} \,.
\end{align}
It gives the leading contribution to $A_n$, which agrees with the prediction of the state integral model: 
\begin{align}
	\begin{split}
		\lim_{n\rightarrow \infty} A_n^{K=b}  &= \log(1+ e^{-5\pi i/3} ) 
		= \frac{1}{6}(3\log 3 + \pi i) \,.
		\\
		\lim_{n\rightarrow \infty} A_n^{K=b^2}  &= \log(1+ 2 e^{-5\pi i/3} + e^{-4\pi i/3}) = \frac{1}{3}(3\log 3 + \pi i) \,.
	\end{split}
\end{align}

To proceed to the 1-loop $(s=1)$ level, we need to improve upon the Gaussian approximation. 
To prepare for a continuum limit, we define 
\begin{align}
	x := \frac{k}{n} \,, 
	\quad 
	0 \le x \le 1 \,.  
\end{align}
A key step is to apply the Euler-Maclaurin formula, including the leading correction term, to approximate $w_{n,k}$ by a continuous function, 
\begin{align}
	\begin{split}
		\frac{w_{n,k}}{w_{n,n/2}}
		\; \approx \; 
		\widetilde{w}_n(x) &:= \exp \left( 2 n \int_{1/2}^x \log[2\sin(\pi y)] dy  + \log\left[\frac{\sin(\pi x)}{\sin(\pi/2)}\right] \right)  \,.
	\end{split}
	\label{wnk-2}
\end{align}
Then, the quantity $E_{n,a}$ can be approximated by the ratio of two integrals,  
\begin{align}
	E_{n,a} \approx \left( \int_0^1 \widetilde{w}_n(x) dx \right)^{-1} 
	\int_0^1 e^{-2\pi i a x} \widetilde{w}_n(x) dx \,.
\end{align}
To evaluate these integrals up to the $(1/n)$ order, we change the variable
\begin{align}
	x = \frac{5}{6} + \frac{1}{\sqrt{n}} t \,, 
\end{align}
and expand the exponents to relevant orders, 
\begin{align}
	\begin{split}
		\widetilde{w}_n(x) &\approx \exp\left( -\sqrt{3}\pi t^2 -\frac{3\sqrt{3}\pi  t + 4\pi^2 t^3 }{3\sqrt{n}} \right) \,, \\
		e^{-2\pi i a x} &\approx e^{-5\pi i a /6 } \exp\left(- \frac{2\pi i a}{\sqrt{n}} t \right) \,.
	\end{split}
\end{align}
Treating the $(1/\sqrt{n})$ terms in the exponents perturbatively, we find 
\begin{align}
	\begin{split}
		A_n^{K_b} &= \frac{1}{6}(3\log 3 + \pi i) + \frac{6 + i \sqrt{3}}{18} \left( \frac{ 2\pi i}{n}  \right) + O\left(\frac{1}{n}\right)^2 \,,
		\\
		A_n^{K_b^2} &= \frac{1}{3}(3\log 3 + \pi i) + \frac{2 + i \sqrt{3}}{3} \left( \frac{ 2\pi i}{n}  \right) + O\left(\frac{1}{n}\right)^2 \,,
	\end{split}
\end{align}
in perfect agreement with \eqref{2-loop-summary} 
to the one-loop order $(s=1)$ up to an additive factor of $i\pi$ in $A_n^{K_b}$ which is due to the $(-1)$ in our definition of \eqref{Jn2}. 

\subsection{Test of the length conjecture - numerical check}

For a numerical check below, we define two sequences from the truncated invariants,
\begin{align}
    B^{K}_n &:= \begin{cases} \sum_{s=0}^{2} \mathcal{W}_s (\hat{O}_{K};S^3\backslash \mathbf{4}_1) \hbar^{s} \bigg{|}_{q=\exp(\frac{2\pi i}n)}   - i \pi ,  & K=K_b
        \quad 
        \\
        \sum_{s=0}^{2} \mathcal{W}_s (\hat{O}_{K};S^3\backslash \mathbf{4}_1) \hbar^{s}   \bigg{|}_{q=\exp(\frac{2\pi i}n)}, &  K=K_b^2\;.
        \end{cases}
\end{align}
On the Jones-polynomial side, we introduce two sequences, 
\begin{align}
    \begin{split}
        A^{K_b}_n  &:= \log \frac{ (-1)\times  J_{n,\tilde{n}=2}(\mathbf{4}_1 \cup K_b;q)}{J_{n}(\mathbf{4}_1 ;q )} \bigg{|}_{q=\exp(\frac{2\pi i}n)} \;,
        \\
        A^{K_b^{ 2}}_n &:= \log \frac{J_{n,\tilde{n}=2}(\mathbf{4}_1 \cup K_b^2;q)}{J_{n}(\mathbf{4}_1 ;q )} \bigg{|}_{q=\exp(\frac{2\pi i}n)} 
        \\
        &= \log \frac{J_{n,\tilde{n}=3}(\mathbf{4}_1 \cup K_b;q)+J_{n}(\mathbf{4}_1 ;q)}{J_{n}(\mathbf{4}_1 ;q )} \bigg{|}_{q=\exp(\frac{2\pi i}n)} \;.
    \end{split}
\end{align}
The numerical values of $A_n^K$ up to $n=30$ are given in Table~\ref{table:numeric}. 
\begin{table}
\begin{align}
    \begin{array}{r|cc|cc}
     n & A_n^{K_b} & C_n^{K_b} & A_n^{K_b^{ 2}} & C_n^{K_b^{ 2}} 
    \\ \hline
       2 & 0.4700 +1.5708i & 3.10 &               &  
    \\ 3 & 0.4345 +1.1336i & 2.74 &-0.6844 +0.9147i & 35.4  
    \\ 4 & 0.4321 +0.9545i & 2.34 &-0.4575 +1.2120i & 46.8
    \\ 5 & 0.4403 +0.8650i & 3.67 &-0.1391 +1.4199i & 49.7
    \\ 6 & 0.4515 +0.8126i & 5.42 & 0.1251 +1.4814i & 49.8
    \\ 7 & 0.4627 +0.7774i & 6.71 & 0.3133 +1.4831i & 48.6
    \\ 8 & 0.4725 +0.7512i & 7.38 & 0.4469 +1.4640i & 46.5
    \\ 9 & 0.4808 +0.7304i & 7.56 & 0.5444 +1.4387i & 44.1
    \\10 & 0.4877 +0.7131i & 7.41 & 0.6176 +1.4126i & 41.5
    \\11 & 0.4935 +0.6983i & 7.06 & 0.6742 +1.3878i & 39.0
    \\12 & 0.4983 +0.6856i & 6.63 & 0.7190 +1.3651i & 36.8
    \\13 & 0.5023 +0.6744i & 6.19 & 0.7552 +1.3446i & 34.9
    \\14 & 0.5058 +0.6647i & 5.78 & 0.7851 +1.3261i & 33.2
    \\15 & 0.5087 +0.6560i & 5.42 & 0.8102 +1.3096i & 31.8
    \\16 & 0.5113 +0.6483i & 5.11 & 0.8316 +1.2947i & 30.6
    \\17 & 0.5136 +0.6414i & 4.85 & 0.8499 +1.2813i & 29.7
    \\18 & 0.5156 +0.6353i & 4.63 & 0.8659 +1.2693i & 28.9
    \\19 & 0.5174 +0.6297i & 4.46 & 0.8800 +1.2583i & 28.2
    \\20 & 0.5190 +0.6246i & 4.31 & 0.8924 +1.2484i & 27.7
    \\21 & 0.5205 +0.6200i & 4.19 & 0.9035 +1.2393i & 27.2
    \\22 & 0.5218 +0.6158i & 4.09 & 0.9135 +1.2310i & 26.8
    \\23 & 0.5230 +0.6119i & 4.01 & 0.9225 +1.2233i & 26.4
    \\24 & 0.5241 +0.6084i & 3.94 & 0.9306 +1.2163i & 26.1
    \\25 & 0.5251 +0.6051i & 3.88 & 0.9381 +1.2098i & 25.9
    \\26 & 0.5260 +0.6021i & 3.83 & 0.9449 +1.2038i & 25.6
    \\27 & 0.5269 +0.5992i & 3.78 & 0.9511 +1.1982i & 25.4
    \\28 & 0.5277 +0.5966i & 3.74 & 0.9569 +1.1929i & 25.2
    \\29 & 0.5284 +0.5942i & 3.71 & 0.9622 +1.1881i & 25.1 
    \\30 & 0.5291 +0.5919i & 3.67 & 0.9672 +1.1835i & 24.9
    \\\infty & \frac{1}{2}\log 3 + \frac{\pi i}{6} &  & \log 3 + \frac{\pi i}{3} &  
    \\       & \approx 0.5493 + 0.5236i & O(1) & \approx 1.0986 + 1.0472i & O(1)
    \end{array}
    \nonumber
\end{align}
\caption{Numerical test of the length conjecture for the $\mathbf{4}_1$ knot with $K_b$ and $K_b^{ 2}$.}
\label{table:numeric}
\end{table}
\begin{figure}[htbp]
	\begin{center}
		\includegraphics[width=0.75\textwidth]{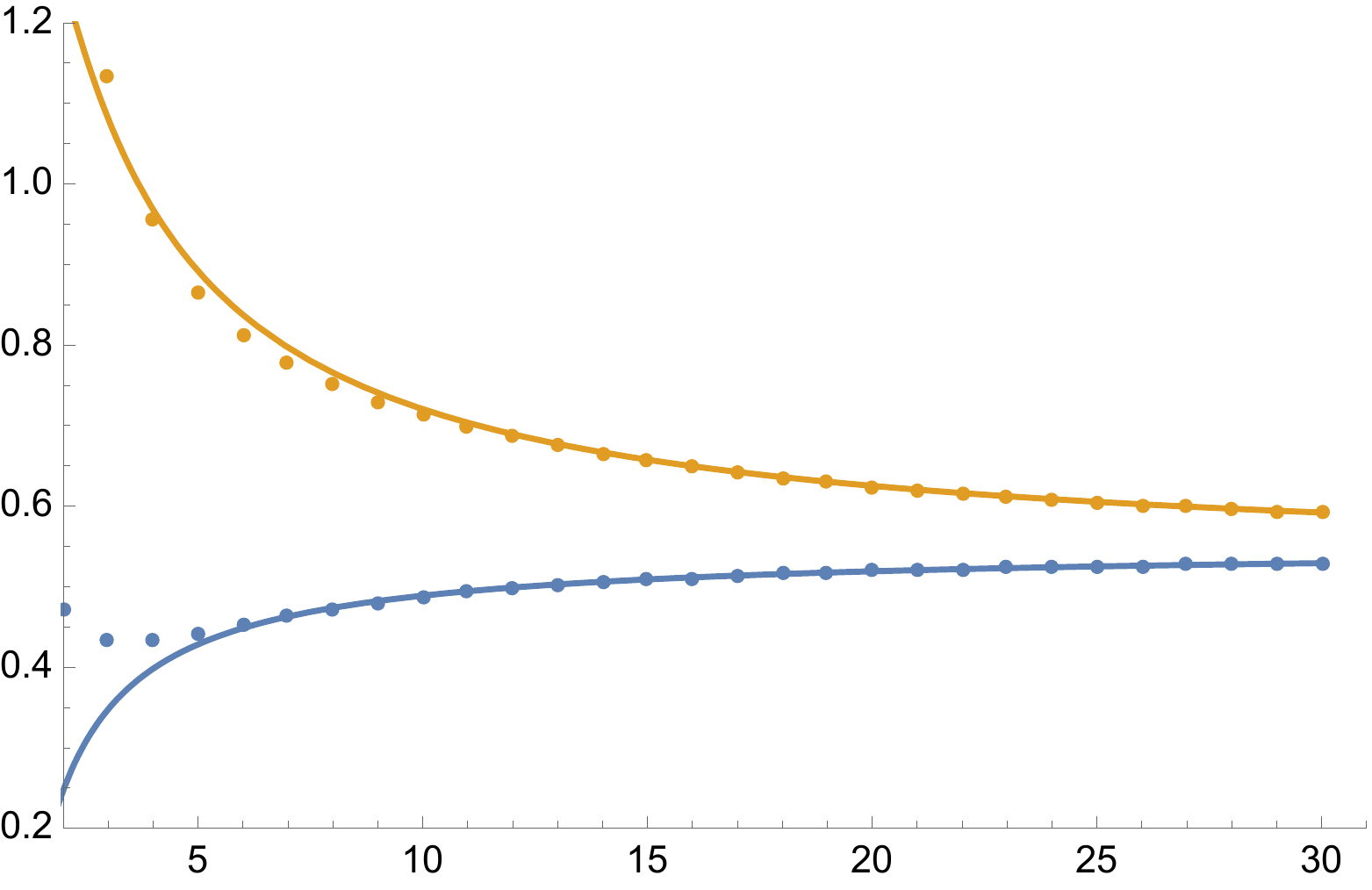} \\
		\includegraphics[width=0.75\textwidth]{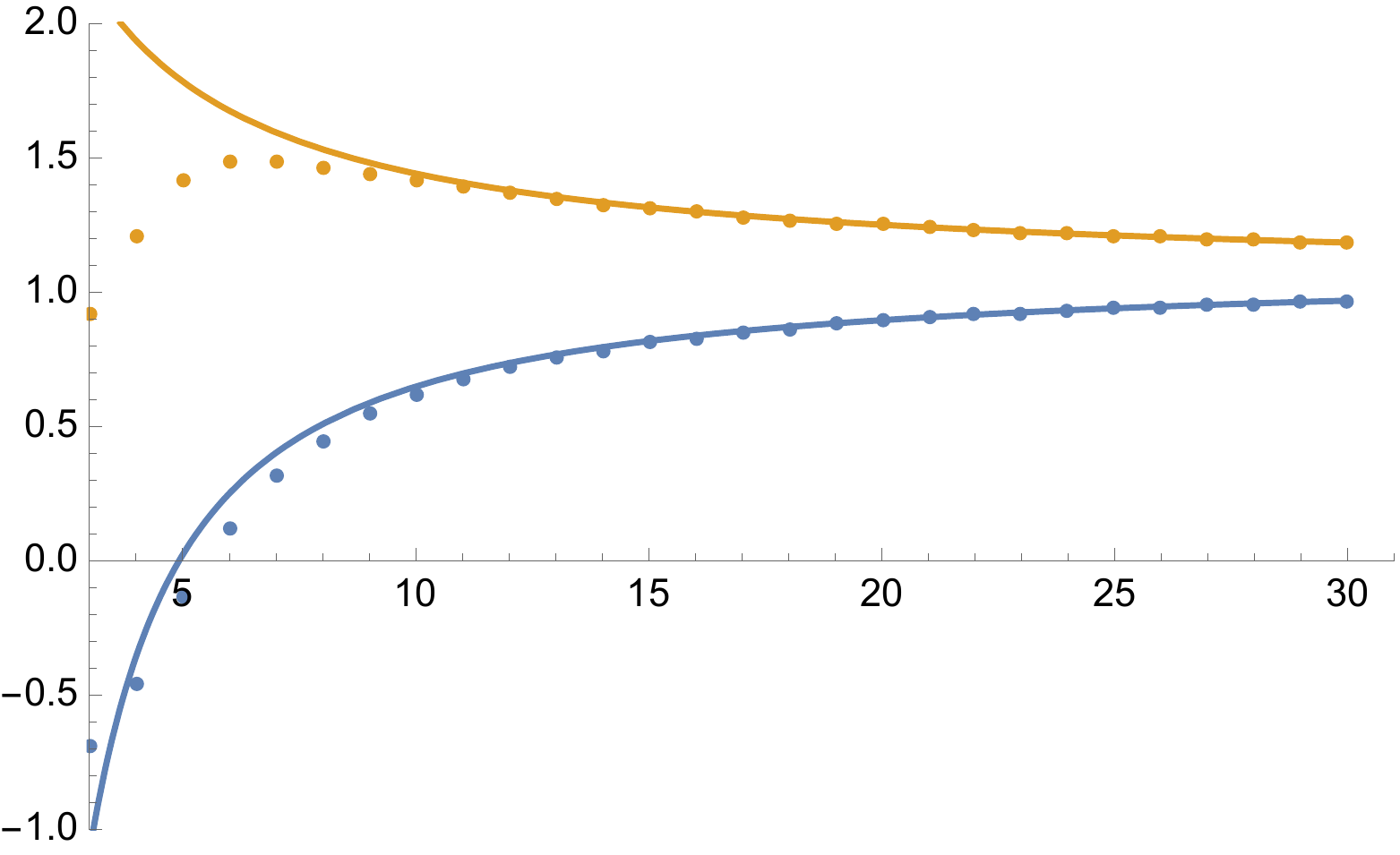}
	\end{center}
	\caption{Numerical test of the length conjecture for $K_b$ (above) and $K_b^{ 2}$ (below). 
	The lower/upper data on the same plot represent the real/imaginary parts, respectively. 
	The dots are the numerical data from $A_n^{K}$, the curves are from the continuous function $B_n^K$.}
	\label{fig:4_1-numerics-1}
\end{figure}

The numerical data for $A_n$ is plotted against the continuous graph of $B_n$ in Figure~\ref{fig:4_1-numerics-1}. 
The two plots appear to converge as $n$ increases. To estimate the error more precisely, 
we define the two-loop error coefficients as  
\begin{align}
    \begin{split}
        C_n^{K} = n^3 | A_n^{K} - B_n^{K} | \,,
        \quad 
        K_b \mbox{ or } K_b^{ 2} \,.
    \end{split}
\end{align}
Their numerical values are also given in Table~\ref{table:numeric}. 
After some fluctuations for small $n$, say $n \le 15$, 
they begin to converge slowly to a fixed $O(1)$ constant. 
This tendency suggests that the error $|A_n - B_n|$ is approximately $O(1/n)^3$. 
So, as far as the numerical experiment is concerned, we have confirmed that the all order length conjecture holds up to two-loop order $(s=2)$.

\section{3D index  with quantum trace map }  \label{sec : index}

We discuss the 3D index with quantum trace map, which we call $\mathcal{I}_{M+\hat{O}} (m,e;q)$ and $\mathcal{I}_{M_{P/Q}+\hat{O}} (q)$, extending the previous works \cite{Dimofte:2011py,garoufalidis20123d,Gang:2018wek}.   

The 3D index for a single tetrahedron is 
\begin{align}
\begin{split}
&\mathcal{I}_{\Delta}(m,e):= \sum_{n=[e]}^{\infty} \frac{(-1)^n q^{\frac{1}2 n (n+1)-(n+ \frac{1}2 e)}}{(q)_n (q)_{n+e}}\;, 
\\
&\textrm{ where }[e]:=\frac{1}2 (|e|-e) \textrm{ and } (q)_n:= (1-q)(1-q^2)\ldots(1-q^{n})\;.
\end{split}
\end{align}
The 3D index for  $M = S^3\backslash \mathcal{K}$ with an insertion on quantum loop operator 
\begin{align}
    \hat{O} = \sum_\a C_\a (q^{1/4}) \prod_{i=1}^k \hat{z}_i^{a_i^{(\a)}} (\hat{z}''_i)^{b_i^{(\a)}}\in \hat{\mathbb{C}}_q [\mathcal{T}]
\end{align}
is given as
\begin{align}
\begin{split}
&\mathcal{I}_{M+\hat{O}} (m,e;q) 
\\
&= \sum_\alpha C_\a \left( \sum_{\{m\}} (-q^{-1/2})^{\langle \nu, \gamma\rangle }  
\left. 
\prod_{i=1}^k \mathcal{I}_\Delta (m_i+b_i^{(\alpha)},e_i-a_i^{(\a)})
q^{\left[a,b;m,e\right]_i^{(\alpha)}} 
\right|_{ \gamma = g^{-1} \cdot \tilde{\gamma}}\right)\;,
\\
&\textrm{with} \quad \{m\} = \{ m_{c_1},\ldots , m_{c_{k-1}} \} \;,
\\
& \gamma := (m_1, \ldots, m_k, e_1, \ldots, e_k)^{T}\;, \quad \tilde{\gamma} = (m,m_{c_1}, \ldots, m_{c_{k-1}}, e,0,\ldots, 0)^T\;,
\\
& \langle \nu, \gamma\rangle := \sum_{i=1}^{k} ((\nu_p)_{i}m_i - \nu_i e_i)\;, 
\quad 
\left[a,b;m,e\right]_i^{(\alpha)} = \frac{1}{2} \left( a_i^{(\alpha)} m_i +b_i^{(\a)}e_i - b_i^{(\a)} a_i^{(\a)} \right) \;.
\end{split}
\end{align}
Here $g\in Sp(2k, \mathbb{Z})$ and $(\vec{\nu}, \vec{\nu}_p) \in \mathbb{Z}^k$ are determined by gluing equations of an ideal triangulation $\mathcal{T}$ of $M$ as given in \eqref{g, nu in gluing eqns}. Under the Dehn filling, $M\rightarrow M_{P/Q}$, with a slope $P/Q$, the 3D index becomes \cite{Gang:2018wek}
\begin{align}
	\begin{split}
		&\mathcal{I}_{ M_{P/Q}+ \hat{O}}
		\\	
		&= \sum_{(m, e)\in \mathbb{Z}^2} \frac{1}2 (-1)^{R m +2 S e} \bigg{(} \delta_{\frac{P}2 m+Qe,0} (q^{\frac{R m+2 S e}2}+ q^{-\frac{R m+2 S e}2})
		\\
		& \qquad \qquad \qquad \qquad    \qquad \qquad \qquad   - \delta_{\frac{P}2 m+Qe,-1}-\delta_{\frac{P}2 m+Qe,1} \bigg{)}  \mathcal{I}_{M+\hat{O}} (m, e)\;,
	\end{split}
\end{align}
Here two integers $(R,S)$ are chosen to satisfy
\begin{align}
	\left(\begin{array}{cc}R & S \\ P & Q\end{array}\right) \in SL(2,\mathbb{Z})\;.
\end{align}
For given $(P,Q)$, two integers $(R,S)$ satisfying the $SL(2,\mathbb{Z})$ relation is not unique but the choice of them does not affect  the 3D index.

\conjecture[]{
	The quantum trace map $\textrm{tr}^{\mathcal{T}}_q$  in the conjecture \ref{main conjecture}   satisfies the following properties}
\begin{align}
	\begin{split}
	& \textrm{For any even link $K \subset M$},
    \\
	&\mathcal{I}_{M+\hat{O}_K} (m,e;q) = \mathcal{I}_{M +\hat{O}} (-m,-e;q)\; \textrm{ and }
	\\
	&\mathcal{I}_{M_{P/Q}+\hat{O}_K} (m,e;q) = 0\;, \; \textrm{ if $M_{P/Q}$ is a Lens space}\;.
	\end{split} \label{conjecture for 3D index}
\end{align}
Not all $\hat{O} \in \hat{\mathbb{C}}_q[\mathcal{T}]$ satisfy the above constraints and it will provide a guideline for determining the quantum trace map. 

\example[]{As an example, consider the case when $M= S^3 \backslash \mathbf{4}_1$. The 3D index $\mathcal{I}_M (m,e;q)$ is
	\begin{align}
		\begin{split}
			&\mathcal{I}_{M=S^3\backslash \mathbf{4}_1} (m,e;q) 
			\\
			&\quad = \sum_{e_c\in \mathbb{Z}}  \mathcal{I}_{\Delta}(e_c-e,e_c)\mathcal{I}_{\Delta}(e_c-m,e_c- e-m)\;.
		\end{split}
	\end{align}
	With an  insertion of $\hat{O}=\hat{y}^\alpha \hat{z}^\beta (\hat{y}'')^\gamma (\hat{z}'')^\delta \in \hat{\mathbb{C}}_q[\mathcal{T}]$, the index becomes
	\begin{align}
		\begin{split}
			\mathcal{I}_{S^3\backslash \mathbf{4}_1+\hat{O}} (m,e;q) 
			&= \sum_{e_c \in \mathbb{Z}} q^{\frac{\alpha (e_c-e)+\beta (e_c-m)+\gamma e_c+\delta (e_c-e-m)-\alpha \gamma - \beta \delta }2}
			\\
			& \quad \;\;  \times \mathcal{I}_\Delta(e_c-e+\gamma,e_c-\alpha) \mathcal{I}_\Delta(e_c-m+\delta,e_c-e-m-\beta)\;.
		\end{split}
	\end{align}
Using the above expression, one can check that all $\hat{O} \in \textrm{tr}^{\mathcal{T}}_q (\mathcal{S}_q^{\rm even}[S^3\backslash \mathbf{4}_1])$ in \eqref{trace map for 41} satisfy the conditions in the conjecture.  
For $\hat{O} =C_1 \hat{y}^{-1} \hat{z}'' + C_2 \hat{z}^{-1} \hat{y}'' +  C_3 \hat{y}^{-1} \hat{z}^{-1}   \in \hat{\mathbb{C}}_q [\mathcal{T}]$, as an example, the  3D index becomes

\begin{align}
	\begin{split}
		&\mathcal{I}_{S^3\backslash \mathbf{4}_1 +\hat{O} } (m,e) 
		\\
		&= \sum_{e_c\in \mathbb{Z}} \big{[} C_1 \mathcal{I}_{\Delta}(e_c-e,e_c+1)\mathcal{I}_{\Delta}(e_c-m+1,e_c- e-m) q^{-\frac{m}2}\;
		\\
		& \quad \quad \quad  +  C_2 \mathcal{I}_{\Delta}(e_c-e+1,e_c)\mathcal{I}_{\Delta}(e_c-m,e_c- e-m+1) q^{\frac{m}2}\;
		\\ 
		&   \quad \quad \quad   +C_3 \mathcal{I}_{\Delta}(e_c-e,e_c+1)\mathcal{I}_{\Delta}(e_c-m,e_c- e-m+1) q^{\frac{m-2e_c +e}2}\;  \big{]}\;,
	\end{split}
\end{align}
which implies (to avoid clutter, we suppress the subscript in $\mathcal{I}_{S^3\backslash \mathbf{4}_1  +\hat{O} }$)
\begin{align}
	\begin{split}	
		\mathcal{I}(0,0) &= C_3 + (-2C_1-2C_2+C_3)q -2(C_1+C_2+C_3)q^2 
		\\
		&\qquad +(2C_1 +2C_2 -5 C_3) q^3+\ldots \,,
		\\
		\mathcal{I}(1,0) &= (C_3-C_1) q^{1/2} - (C_1+C_2+C_3)q^{3/2}+ (2C_1-C_2 -3C_3)q^{5/2} +\ldots \,,
		\\
		\mathcal{I} (-1,0) 
		&= (C_3-C_2 )q^{1/2}-(C_1+C_2+C_3)q^{3/2}+(2C_2-C_1 -3C_3)q^{5/2}+\ldots \,,
		\\
		\mathcal{I} (1,1) 
		&=(C_3-C_1)-C_3 q +(C_1 -2C_2-2C_3)q^2+(4C_1-3C_3)q^3+\ldots \,,
		\\
		\mathcal{I} (-1,-1) 
		&=(C_3-C_2)-C_3 q +(C_2 -2C_1-2C_3)q^2+(4C_2-3C_3)q^3+\ldots \,. \nonumber
	\end{split}
\end{align}
For the index to be invariant under the Weyl $\mathbb{Z}_2$ symmetry, $(m,e) \leftrightarrow (-m,-e)$,  in  the conjecture, 
the quantum loop operator $\hat{O} =C_1 \hat{y}^{-1} \hat{z}'' + C_2 \hat{z}^{-1} \hat{y}'' +  C_3 \hat{y}^{-1} \hat{z}^{-1} $  should satisfy
\begin{align}
C_1 = C_2\;.
\end{align}
So we expect that only  the $\hat{O}$ with $C_1=C_2$ can appear as image of the quantum trace map. For example, from \eqref{trace map for 41}, 
\begin{align}
\begin{split}
    \textrm{tr}_q^{\mathcal{T}}(K_b) &= (\hat{y}^{-1}+\hat{z}^{-1}-\hat{y}^{-1}\hat{z}^{-1}) 
    \\
    &\sim (\hat{y}^{-1} (\hat{z}''+\hat{z}^{-1})+\hat{z}^{-1}(1-\hat{y}^{-1})) 
    \\
    &\sim (\hat{y}^{-1}(\hat{z}''+\hat{z}^{-1})+\hat{z}^{-1}\hat{y}'')   = \hat{O}|_{C_1=C_2=C_3=-1}
\end{split}
\end{align} 
satisfies the condition.  
}

\section{Quantum trace map in complex Chern-Simons theory} 
In this section, we give a brief review on general aspect of $SL(2,\mathbb{C})$ Chern-Simons theory and its relation the volume conjecture. Refer to \cite{gukov2005three,dimofte2009exact,Witten:2010cx} for details. By interpreting the quantum trace operator as holomorphic Wilson loop in the complex Chern-Simons theory, we give a physical derivation of the length conjecture. 

The action of $SL(2,\mathbb{C})_{(k,\sigma)}$ Chern-Simons theory is given as follows
\begin{align}
S_{(k,\sigma)}[\CA, \overline{\CA};M] = \frac{k + \sigma}{8\pi } \int \textrm{tr} \left(\CA \wedge d \CA+\frac{2}3 \CA^3\right) + \frac{k - \sigma}{8\pi } \int \textrm{tr} \left(\overline{\CA} \wedge d \overline{\CA}+\frac{2}3 \overline{\CA}^3\right)\;. \nonumber
\end{align}
The $k\in \mathbb{Z}$ should be quantized for gauge invariance while $\sigma$ can be either real ($\sigma \in \mathbb{R}$) or purely imaginary ($\sigma \in i \mathbb{R}$). The state-integral  and the 3D index for a knot complement $M= S^3\backslash \mathcal{K}$ compute the partition function $Z$ of the complex Chern-Simons theory with $k=1$ and $k=0$ respectively \cite{Dimofte:2011py,Dimofte:2014zga}: 
\begin{align}
	\begin{split}
Z_{\hbar=2\pi i b^2} (M; X_{\bf m }) &=  Z \bigg{[}SL(2,\mathbb{C})_{k, \sigma } \textrm{ CS theory on } M \textrm{ with }k=1, \sigma = \frac{1-b^2}{1+b^2} \bigg{]}\;
\\
&=\int \frac{[D\CA][D\overline{\CA}]}{(\rm gauge)}\bigg{|}_{b.c} \exp \left( i S_{(k, \sigma)} [\CA, \overline{\CA};M] \right)\big{|}_{k=1, \sigma = \frac{1-b^2}{1+b^2}}\;,
\\
\mathcal{I}^{\rm fugacity}_{M} (m,u;q) &=  Z \bigg{[}SL(2,\mathbb{C})_{k, \sigma } \textrm{ CS theory on } M \textrm{ with }k=0, \sigma = \frac{4\pi i }{\log q} \bigg{]}\;
\\
&=\int \frac{[D\CA][D\overline{\CA}]}{(\rm gauge)}\bigg{|}_{b.c} \exp \left( i S_{(k, \sigma)} [\CA, \overline{\CA};M] \right)\big{|}_{k=0, \sigma = \frac{4\pi i}{\log q}}\;.
\end{split} \label{Z and I in path-integral}
\end{align}
Here $\mathcal{I}^{\rm fugacity}_M(m,u;q)$ is the 3D index in fugacity basis
\begin{align}
\mathcal{I}^{\rm fugacity}_M (m,u;q) = \sum_{e \in \mathbb{Z}} \mathcal{I}_M (m,e;q) u^e\;.
\end{align}
The boundary conditions are
\begin{align}
\begin{split}
&P \exp \left(\oint_{O_{\rm m }} \CA \right) = \begin{pmatrix}
	e^{X/2} & * \\
	0 & e^{-X/2}
\end{pmatrix}, \quad P \exp \left(\oint_{O_{\rm m }} \overline{\CA} \right) = \begin{pmatrix}
e^{\tilde{X}/2} & * \\
0 & e^{-\tilde{X}/2}
\end{pmatrix}
\\
& \textrm{where}
\\
&  (e^X, e^{\tilde{X}}) = \begin{cases} (e^{X_{\bf m}}, e^{\frac{X_{\bf m}}{b^2}}) & \textrm{for } Z_{\hbar =2\pi i b^2} (M;X_{\bf m})
	\\
	(q^{\frac{m}2}u, q^{\frac{m}2}u^{-1}) & \textrm{for } \mathcal{I}^{\rm fugacity}_M(m,u;q)
 \end{cases}
\end{split}\label{bc for CS ptn}
\end{align}
The partition functions can be written in the following factorization form  \cite{Beem:2012mb}
\begin{align}
	\begin{split}
Z_{\hbar  = 2\pi i b^2}(M;X_{\bf m }) &=\frac{1}2 \sum_{\rho} B_M^{\rho} (q;e^{X}) B_M^{\rho}(\tilde{q}; e^{\tilde{X}}) \textrm{ with }  (q,\tilde{q})=(e^{2\pi i b^2},e^{\frac{2\pi i} {b^2}})\;,
\\
\mathcal{I}^{\rm fugacity}(m,u;q)&=\frac{1}2 \sum_{\rho} B_M^{\rho} (q;X) B_M^{\rho}(\tilde{q}; \tilde{X}) \textrm{ with }  (q,\tilde{q})=(q,q^{-1})\;. 
\end{split}\label{factorization of Z and I}
\end{align}
$(e^{X},e^{\tilde{X}})$ here are identical to that of \eqref{bc for CS ptn}. 
$B_M^{\rho}(q;e^X)$ is so-called holomorphic block labelled by an (adj)-irreducible $SL(2,\mathbb{C})$ flat connection $\CA_\rho$ satisfying the boundary condition in \eqref{bc for CS ptn}.\footnote{A $SL(2,\mathbb{C})$ flat connection $\rho \in \textrm{Hom}[\pi_M \rightarrow SL(2,\mathbb{C})]$ can be alternatively described by a $SL(2,\mathbb{C})$ gauge connection $\mathcal{A}_{\rho}$ with vanishing curvature, i.e. $d\mathcal{A}_\rho + \mathcal{A}_\rho \wedge \mathcal{A}_\rho =0$. } It can be defined as following path-integral
\begin{align}
B_M^{\rho}(q;e^{X}) = \int_{\Gamma^\rho} [D\CA]  e^{-\frac{1}{2\hbar }\int_M \textrm{tr}(\CA \wedge d\CA +\frac{2}3 \CA^3) }\; \textrm{ with } q=e^\hbar\;. \label{path-integral of B}
\end{align}
Here $\Gamma^{\rho}$ denotes the Lefschetz thimble in the functional space of $\CA$ associated with the flat connection $\CA_\rho$ \cite{Witten:2010cx}. 

The $Z_{\hbar}(M, X_{\bf m})$ and $B_M(q=e^{\hbar};e^{X})$ have the same asymptotic expansion in the limit $\hbar \rightarrow 0$:
\begin{align}
B^{\rho_\a}_M (q=e^{\hbar };e^{X}) \xrightarrow{\quad \hbar \rightarrow 0 \quad } \exp \left( \sum_{n=0}^{\infty} S_n^{(\alpha)}(M) \hbar^{n-1} \right)\bigg{|}_{X_{\bf m}=X} \label{B-and-Z}
\end{align}
where the $S_n^{(\alpha)}$ is the perturbative expansion of the state-integral around the saddle point $\vec{Z}^{(\alpha)}$ as defined in \eqref{perturbative Sna}. For each irreducible flat connection $\rho_\alpha$, there is an associated saddle point $\vec{Z}^{(\alpha)}$ in the state-integral model. More precisely, the map is generically in 2-to-1 since two flat connections $\rho_\alpha$ and $\rho_{\tilde{\alpha}} $  related to each other by the action of $H^1 (M, \mathbb{Z}_2)= \mathbb{Z}_2$ correspond to the same saddle point $\vec{Z}^{(\alpha)}$. 
In the asymptotic limit $\hbar \rightarrow 0$, the  $B^{\rho}_M (\tilde{q}; e^{\tilde{X}})$ becomes trivially $1$ since $\tilde{q} = e^{-\frac{4\pi^2}\hbar}$ goes to 0 in the limit and $B^{(\rho)}(q)$ becomes an infinite power series in $q^{1/2}$ starting with $1+O(q^{1/2})$ as $q\rightarrow 0$.

The quantum trace map $\hat{O}_K$ for a knot $K \subset M$ corresponds to the following holomorphic Wilson loop in the CS theory
\begin{align}
W_{K} (\CA) =(-1)\times \textrm{Tr} \left( P \exp \left( \oint_K \CA\right) \right)\;,
\end{align}
and the $Z_\hbar (M+\hat{O}_K)$ and $\mathcal{I}_{M+\hat{O}_K}$ are given by following path-integrals
\begin{align}
\begin{split}
Z_\hbar (M+\hat{O}_K;X_{\bf m})  &=(-1)\times \int \frac{[D\CA][D\overline{\CA}]}{(\rm gauge)} e^{ i S_{(k, \sigma)} [\CA, \overline{\CA};M]}W_K(\CA)\big{|}_{k=1, \sigma = \frac{1-b^2}{1+b^2}}\;,
\\
\mathcal{I}^{\rm fugacity}_{M+\hat{O}_K} (m,u;q)  &=(-1)\times \int \frac{[D\CA][D\overline{\CA}]}{(\rm gauge)} e^{ i S_{(k, \sigma)} [\CA, \overline{\CA};M]}W_K(\CA)\big{|}_{k=0, \sigma =\frac{4\pi i}{\log q}}\;.
\end{split}
\end{align}
With the insertion of  loop operator, we expect that factorization should be modified as follows
\begin{align}
	\begin{split}
		Z_{\hbar  }(M+\hat{O};X_{\bf m }) &=\frac{1}2 \sum_{\rho} B_{M+\hat{O}}^{\rho} (q;e^X) B_M^{\rho}(\tilde{q}; e^{\tilde{X}})\;,
		\\
		\mathcal{I}^{\rm fugacity}(m,u;q)&=\frac{1}2 \sum_{\rho} B_{M+\hat{O}}^{\rho} (q;e^X) B_M^{\rho}(\tilde{q}; e^{\tilde{X}})\;.
	\end{split}
\end{align}
Here we define
\begin{align}
	B_{M+\hat{O}_K}^{\rho}(q;X) = (-1)\times \int_{\Gamma^\rho} [D\CA]  e^{\frac{i}{2\hbar }\int_M \textrm{tr}(\CA \wedge d\CA +\frac{2}3 \CA^3) } W_K (\mathcal{A})\;. \label{path-integral of B-2}
\end{align}
In the same argument used in \eqref{B-and-Z},  the $Z_{\hbar}(M+\hat{O}_K, X_{\bf m})$ and $B_{M+\hat{O}_K}(q=e^{\hbar};e^{X})$ is expected to share the same asymptotic expansion in the limit $\hbar \rightarrow 0$:
\begin{align}
	B^{\rho_\a}_{M+\hat{O}_K} (q=e^{\hbar };e^{X}) \xrightarrow{\quad \hbar \rightarrow 0 \quad } \exp \left( \sum_{n=0}^{\infty} S_n^{(\alpha)}(\hat{O}_K;M) \hbar^{n-1} \right)\bigg{|}_{X_{\bf m}=X}\;. \label{B-and-Z-2}
\end{align}
where the $S_n^{(\alpha)}$ is the perturbative expansion of the state-integral around the saddle point $\vec{Z}^{(\alpha)}$ as defined in \eqref{perturbative Sna}.

Volume conjecture for a hyperbolic knot $\mathcal{K}$ relates an asymptotic expansion of Jones polynomial to an asymptotic expansion of the $B^{\rm (hyp)}_{S^3\backslash \mathcal{K}}(q,X)$ as follows
\begin{align}
\begin{split}
&\bigg{(}\textrm{asymptotic expansion of $\frac{J_n\left(\mathcal{K}; q=\exp (\frac{2\pi i }{k})\right)}{J_n\left(\bigcirc; q=\exp (\frac{2\pi i }{k})\right)}$ in the limit $k=n \rightarrow \infty $} \bigg{)}
\\
& \simeq \bigg{(}\textrm{asymptotic expansion of $B^{\rho_{(\rm hyp) }}_{S^3\backslash \mathcal{K}} (q=e^{\hbar},X=0)$ in the limit $\hbar \rightarrow 0 $} \bigg{)}\;,
\\
& \textrm{with an identification $\hbar = \frac{2\pi i }k$}\;. \label{B-and-J}
\end{split}
\end{align}
Here $\bigcirc$ denote the unknot in $S^3$ and $\simeq $ means the same asymptotic expansion modulo an overall factor of the form $\frac{C}{k^{3/2}}$.
The equivalence of two asymptotic expansions can be understood by comparing the path-integral of the Jones polynomial  in \eqref{path-integral of jones 2} and the $B^{\rho}_M$  in \eqref{path-integral of B} modulo some subtleties.  The most subtle part is that how the path-integral of $SU(2)$ gauge field  $a$ can be related to the path-integral of $SL(2,\mathbb{C})=SU(2)_{\mathbb{C}}$ gauge field $\mathcal{A}$ along the $\Gamma^{\rm (hyp)}$ in the asymptotic limit. To understand it,   let us consider a finite dimensional integral $\int_{\mathbb{R}} dx e^{i k (\frac{1}3x^3+x)}$ in the limit $k\rightarrow \infty$ as an analogy. Although it is integration over real axis, it gets contribution from the saddle point at $x=-i$ in the asymptotic limit. So, the appearance of complex flat connection in the asymptotic expansion of $SU(2)$ Chern-Simons theory is not so strange. But unlike in the finite-dimensional analogy, the contribution from the flat connection $\rho_{\rm (hyp)}$ is exponentially growing in the asymptotic limit. This issue has been addressed in \cite{Witten:2010cx} and we will not address the subtle issue here. 

Our length conjecture can be understood in a similar way. For that, we consider the Jones polynomial $J_{n,\tilde{n}=2} (\mathcal{K} \cup K;q)$. We consider the case when the $K$ is  a knot, i.e. $\sharp(K)=1$, for simplicity. The Jones polynomial can be represented by following  path-integral 

\begin{align}
	\begin{split}
	&J_{n,\tilde{n}=2}\left(\mathcal{K}\cup K ;q= \exp (\frac{2\pi i }{k}) \right) 
	\\
	&=(-1)\times \frac{ \int \frac{[da]}{(\rm gauge)} \bigg{|}_{\textrm{b.c. in \eqref{b.c. for Jones}}} \exp \left(\frac{i(k-2)}{4\pi} \int_{S^3 \backslash \mathcal{K}} \textrm{Tr}(a\wedge da - \frac{2}3 a^3)\right) W_K (a) }{  \int \frac{[da]}{(\rm gauge)} \exp \left(\frac{i(k-2)}{4\pi} \int_{S^3} \textrm{Tr}(a\wedge da - \frac{2}3 a^3)\right)  }\;. \label{path-integral of jones 3}
	\end{split}
\end{align} 
Then, from comparison with the path-integral in \eqref{path-integral of B-2}, one naturally expects that
\begin{align}
	\begin{split}
		&\bigg{(}\textrm{asymptotic expansion of $\frac{J_{n,\tilde{n}=2}\left(\mathcal{K}\cup K; q=\exp (\frac{2\pi i }{k})\right)}{J_n\left(\bigcirc; q=\exp (\frac{2\pi i }{k})\right)}$ in the limit $k=n \rightarrow \infty $} \bigg{)}
		\\
		& \simeq \bigg{(}\textrm{asymptotic expansion of $ B^{\rho_{(\rm hyp)} }_{S^3\backslash \mathcal{K}+\hat{O}_K} (q=e^{\hbar},X=0)$ in the limit $\hbar \rightarrow 0 $} \bigg{)}\;,
		\\
		& \textrm{with an identification $\hbar = \frac{2\pi i }k$}\;.  \label{B-and-J-2}
	\end{split}
\end{align}
Combining \eqref{B-and-J},  \eqref{B-and-J-2} with  \eqref{B-and-Z},  \eqref{B-and-Z-2}, one derives the length conjecture in \eqref{conjecture 2} where the $\mathcal{Z}^{\rm (hyp)}_s (\hat{O}_K;M)$ is given in \eqref{perturbative Zs}.

Proposed properties of $\mathcal{I}_{M+\hat{O}} (m,e;q)$ or $\mathcal{I}_{M_{P/Q}+\hat{O}}$ in \eqref{conjecture for 3D index} simply follows from the path-integral expression \eqref{Z and I in path-integral} or the factorization \eqref{factorization of Z and I}. The invariance of the 3D index under $(m,e) \leftrightarrow (-m,-e)$ comes from the Weyl $\mathbb{Z}_2$ invariance, $X_{\bf m} \leftrightarrow -X_{\bf m}$, in the path-integral. For $M_{P/Q} =(\textrm{Lens space})$, there is no irreducible flat connection $\rho$ on $M_{P/Q}$  and thus we expect the index vanishes from the factorization.

\section*{Acknowledgements}
The work of 
DG is supported in part by the National Research Foundation of Korea grant NRF-2021R1G1A1095318 and by Creative-Pioneering Researchers Program through Seoul National University. 
The work of SL is supported in part by the National Research Foundation of Korea grant NRF-2019R1A2C2084608. MR acknowledges
support from the National Key Research and Development Program of China, grant No. 2020YFA0713000, and the Research Fund for International Young Scientists, NSFC grant No. 11950410500.  

\newpage
\appendix
\section{Gluing equations of an ideal triangulation} \label{gluing}
From an ideal triangulation $\mathcal{T}$ of a knot complement $M=S^3\backslash \mathcal{K}$, we have following gluing equations for the logarithmic edge parameters, $(e^{Z},e^{Z'},e^{Z''})=(z,z'',z'')$, of tetrahedra (refer to, e.g., \cite{Dimofte:2012qj} for details)
\begin{align}
\begin{split}
&C_{I=1,\ldots, |\mathcal{T}|} = \sum_{i=1}^{|\mathcal{T}|} F_{Ii} Z_i + G_{Ii}Z_i' + H_{Ii}Z_i'' - (2\pi i  +\hbar )\textrm{ (internal edges)}\;,
\\
& \mathfrak{M} = \sum_{i=1}^{|\mathcal{T}|} U_{i} Z_i + V_i Z_i' + W_i Z_i''  \textrm{ (meridian)}\;,
\\
& \mathfrak{L} = \sum_{i=1}^{|\mathcal{T}|} P_{i} Z_i + Q_i Z_i' + R_i Z_i''   \textrm{ (longitude)}\;.
\end{split}
\end{align}
Here $F_{Ii}, G_{Ii}, H_{Ii} \in \{0,1,2\}$ while $U_i, V_i, W_i, P_i, Q_i, R_i \in \mathbb{Z}$.  
The logarithmic edge parameters $(Z_i, Z_i',Z_i'')$ satisfy  following linear relation
\begin{align}
Z_i +Z_i'+Z_i'' = i  \pi +\frac{\hbar}2\;.
\end{align}
Then, the gluing equation variety $\chi[\mathcal{T}]$ is given by
\begin{align}
\begin{split}
    	\chi[\mathcal{T} ] = \big{\{} (z_i, z_i'')_{i=1}^{|\mathcal{T}|} &\;:\;   z_i^{-1}+z''_i-1=0, \; 
    	\\
    	&e^{C_I} := \prod_{i=1}^{|\mathcal{T}|} (-1)^{G_{Ii}}z_i^{F_{Ii}-G_{Ii}} (z_i'')^{H_{Ii}-G_{Ii}}=1 \big{\}}\;.
\end{split}
\end{align} 
We define $e^{\hat{C}_I}, e^{\hat{\mathfrak{M}}}$ and $e^{-\hat{\mathfrak{M}}} \in \mathbb{C}[q^{\pm 1/4}][\hat{z}^{\pm 1}_i, \hat{z}''^{\pm 1}_i]/\langle \hat{z}_i \hat{z}''_j -q^{-\delta{ij}} \hat{z}''_j \hat{z}_i\rangle$ as
\begin{align}
&e^{\hat{C}_I} : = q^{-1}\prod_{i=1}^{|\mathcal{T}|}(-q^{\frac{1}2})^{G_{Ii}} q^{\frac{1}2 (F_{Ii}-G_{Ii})(H_{Ii}-G_{Ii})} \hat{z}_i^{F_{Ii}-G_{Ii}} (\hat{z}''_i)^{H_{Ii}-G_{Ii}}\;, \label{exponentiated C}
\\
\begin{split}
&e^{\hat{\mathfrak{M}}} : = \prod_{i=1}^{|\mathcal{T}|}(-q^{\frac{1}2})^{Q_i } q^{\frac{1}2 (P_i -Q_i)(R_i-Q_i)} \hat{z}_i^{P_i-Q_i} (\hat{z}''_i)^{R_i - Q_i}\;, 
\\
&e^{-\hat{\mathfrak{M}}} : = \prod_{i=1}^{|\mathcal{T}|}(-q^{-\frac{1}2})^{Q_i } q^{\frac{1}2 (P_i -Q_i)(R_i-Q_i)} \hat{z}_i^{Q_i-P_i} (\hat{z}''_i)^{Q_i-R_i}\;.
\label{exponentiated M}
\end{split}
\end{align}
They all mutually commute due to a symplectic structures in  gluing equations \cite{neumann1985volumes}
\begin{align}
[e^{\hat{C}_I}, e^{\hat{C}_J}] = [e^{\pm \hat{\mathfrak{M}}}, e^{\hat{C}_I}] =0\;.
\end{align}

As a concrete example, we collect some information on the figure-8 knot complement. 
%
The fundamental group of the knot-complement is (see Figure \ref{fig: figure8-fundamental})
\begin{align}
    \pi_1(S^3\backslash \mathbf{4}_1) = \langle a,b,c \,: \, ac^{-1}b a^{-1}c = bc^{-1}b^{-1}a=1 \rangle \,. 
    \label{pi1-presentation}
\end{align}
It contains the peripheral subgroup $\mathbb{Z}\times \mathbb{Z}$ to be identified as 
the fundamental group of the boundary torus, 
\begin{align}
    \pi_1(\partial(S^3\backslash \mathbf{4}_1)) 
    = \pi_1(\mathbb{T}^2) = \mathbb{Z}\times \mathbb{Z} 
    = \langle \mathbf{m}, \mathbf{l} \rangle 
    \subset \pi_1(S^3\backslash \mathbf{4}_1)  \,. 
\end{align}
We choose $(\mathbf{m}, \mathbf{l})$ canonically  to be the meridian and longitude. 
Then the embedding $i: \pi_1(\partial(S^3\backslash \mathbf{4}_1)) \rightarrow \pi_1(S^3\backslash \mathbf{4}_1)$ 
is given by 
\begin{align}
    i(\mathbf{m}) = a, \quad i(\mathbf{l}) = ac^{-1}bca^{-1}b^{-1} \,.
\end{align}
The simplest ideal triangulation $\mathcal{T}$ of the knot complement consists of two tetrahedra. The edge variables, 
\begin{align}
    (y,y',y'') = (e^Y,e^{Y'} , e^{Y''}) \,,
    \quad 
    (z,z',z'') = (e^Z,e^{Z'} , e^{Z''}) \,,
\end{align}
are subject to the usual conditions, 
\begin{align}
    y + \frac{1}{y'} = 1\,,
    \quad 
    y' + \frac{1}{y''} = 1\,,
    \quad
    y'' + \frac{1}{y} = 1\,,
    \quad 
    y y' y'' = -1, 
\end{align}
and similar ones for $(z,z',z'')$. 
The two internal edges require that 
\begin{align}
\begin{split}
    C_1 &= Y'+ 2Y+Z'+2Z -2\pi i = 0 \,,
    \\
    C_2 &= Y' + 2Y'' + Z' + 2Z'' - 2\pi i = 0\,,
\end{split}
\end{align}
which implies $yz = y'' z''$. In summary, the gluing equation variety $\chi[\mathcal{T}]$ is
\begin{align}
\chi[\mathcal{T}] = \{ (y, z, y'', z'') \;:\; y^{-1}+y''-1=0, z^{-1}+z''-1=0, yz =y'' z'' \}\;. \label{gluing equation variety of 41}
\end{align}
For each $(y,z,y'',z'')\in \chi[\mathcal{T}]$, one can assign a $PSL(2,\mathbb{C})$-representation $\rho$ as follows
%
\begin{align}
\begin{split}
    \rho(a) &=\bigg{[} \frac{1}{\sqrt{y''z}} 
    \begin{pmatrix}
    y''  & 0 \\ -y^{-1} & z
    \end{pmatrix} \bigg{]} \,,
    \\
    \rho(b) &= \bigg{[} 
    \frac{1}{\sqrt{y'z'}}
     \begin{pmatrix}
    z' & -z' \\ y'z' & y' (1-z')
    \end{pmatrix} \bigg{]}  \,,
    \\
    \rho(c) &= \bigg{[} 
         \frac{1}{\sqrt{yz''}} \begin{pmatrix}
    y - z^{-1} & 1-y \\ -z^{-1} & 1
    \end{pmatrix} \bigg{]} \,. \label{hol matrix}
\end{split}
\end{align}
Here we regard $PSL(2,\mathbb{C}) = SL(2,\mathbb{C})/\mathbb{Z}_2$ and $[\ldots]$ is the equivalence class under the $\mathbb{Z}_2 = \{\pm 1\}$. 
These are consistent with the two relations in \eqref{pi1-presentation} and the boundary holonomies deduced from the triangulation, 
\begin{align}
\begin{split}
    \rho(\mathbf{m}) &= \rho(a) = \bigg{[}
    \begin{pmatrix}
    e^{\mathfrak{M}/2} & 0 \\ * & e^{-\mathfrak{M}/2}
    \end{pmatrix} \bigg{]} \,, \quad \mathfrak{M}= Y''-Z \,,
    \\
    \rho(\mathbf{l}) &= \rho(ac^{-1}b ca^{-1}b^{-1}) =  \bigg{[}
    \begin{pmatrix}
    e^{\mathfrak{L}} & 0 \\ * & e^{-\mathfrak{L}}
    \end{pmatrix}  \bigg{]}\,, \quad \mathfrak{L} = Z-Z'' \,.
\end{split}
\end{align}

\section{Colored Jones polynomial for framed link } %
\label{Appendix : Jones polynomial}

Let  $K = \cup_{I=1}^{\sharp(K)} K_I$ be a framed link  inside 3-sphere $S^3$. 
Here we provide the definitions of  colored Jones polynomial $J_{\vec{n}} (K;q)$ using the Kauffman bracket Skein module $\mathcal{S}_q[S^3]$. 
The subscript $\vec{n}  = (n_1, \ldots, n_{\sharp(K)})$ with $n_i \in \mathbb{Z}_{\geq 1}$ represents  {\it colors} of each components of the link. 

When $n_i=2$ for all $i=1,\ldots, \sharp(K)$,  the invariant is called just Jones polynomial and will be simply denoted by $J$ without subscript. The  polynomial $J(K;q)$ is defined as follows
\begin{align}
	J(K;q) :=\frac{Y_{K}}{Y_{\emptyset }} \in \mathbb{Z}[q^{1/4},q^{-1/4}]\;.
\end{align}
Here $Y_K$ is considered to be an element of $\mathcal{S}_q[S^3]$. The definition makes sense since the Skein module $\mathcal{S}_q[S^3]$ is one-dimensional module spanned by $Y_{\emptyset }$. The invariant is slightly different from the conventional Jones polynomial, $J^{\rm oriented}(K;q)$, which is an invariant for oriented (unframed) link. The two invariants are related to each other in the  following way 
%
\begin{align}
J^{\rm oriented}(K;q) :=(-1)^{\sharp(K)}(-q^{3/4})^{ w(K)} J(K;q)
\end{align}
The  sign factor is  introduced in order for  $J^{\rm oriented}(K=\bigcirc^n;q)$ to be $(q^{1/2}+q^{-1/2})^n$.
Here $w(K)$ is the writhe of the framed link
\begin{align}
\begin{split}
w(K) :=\sum_{p: \textrm{crossings in $K$}} \epsilon (p)\;, \quad \epsilon (p) := \begin{cases}
+1, \quad \textrm{$p$ is a positive crossing}\\
-1, \quad \textrm{$p$ is a negative crossing}
\end{cases}
\end{split}
\end{align}
%
The writhe is an oriented framed link invariant. Since the factor $(-q^{3/4})^{w(K)}$ exactly cancels the  $(-q^{3/4})^{\pm 1}$ factor  in $Y_K$ under the 1st Reidemeister move, see Figure \ref{fig: Reidemeister move 1}, the  $J^{\rm oriented}$ is an oriented unframed invariant.  
%
%

To generalize the definition to general colors $\vec{n}$, we introduce the symmetric $k$-product of  a framed knot $K$
\begin{align}
	\begin{split}
		&\textrm{Sym}^{\otimes k=0} K = \emptyset \;, \quad \textrm{Sym}^{\otimes k=1} K = K\;, \quad \textrm{Sym}^{\otimes k=2} K = K^{2} - \emptyset;, \quad
		\\
		&\textrm{Recursively, } \textrm{Sym}^{\otimes k} K :=(\textrm{Sym}^{\otimes k-1} K)  K -(\textrm{Sym}^{\otimes k-2} K)\;. \label{symmetric cabling}
	\end{split}
\end{align}
Here $K^n$ denote the $n$-cabling of a framed knot $K$, a framed link obtained by displacing $n$ parallel copies of $K$ along the normal direction of framing. Then, we define the  framed colored Jones polynomial for a link $K = \bigcup_{I=1}^{\sharp(K)} K_I$  as follows
\begin{align}
J_{\vec{n}}(K;q) :=J(\bigcup_{I=1}^{\sharp(K)}\textrm{Sym}^{\otimes (n_I-1)}K_I;q) = \frac{Y_{\bigcup_{I=1}^{\sharp(K)}\textrm{Sym}^{\otimes (n_I-1)}K_I}}{Y_{\emptyset}} \;.
\end{align}
The framed colored Jones polynomial have following path-integral representation $(k>2)$ \cite{witten1989quantum}
\begin{align}
\begin{split}
&J_{\vec{n}}\left(K;q= q= e^{2\pi i/k} \right) 
\\
&=\prod_{I=1}^{\sharp(K)}(-1)^{n_I-1}\frac{ \int \frac{[da]}{(\rm gauge)} \exp \left(\frac{i(k-2)}{4\pi} \int_{S^3} \textrm{CS}(a)\right) \prod_{I=1}^{\sharp(K)}  \textrm{tr}_{R_n}
\left(P\exp (\oint_{K_I} a)\right)}{  \int \frac{[da]}{(\rm gauge)} \exp \left(\frac{i(k-2)}{4\pi} \int_{S^3} \textrm{CS}(a)  \right)}\;,
\\
&\textrm{where }\textrm{CS}(a) = \textrm{Tr}\left(a\wedge da - \frac{2}3 a^3\right) \;.
\label{path-integral of jones}
\end{split}
\end{align} 
Here $\int \frac{[da]}{(\rm gauge)}$ means integral over gauge-equivalence classes of $SU(2)$ connections on $S^3$. $P\exp (\oint_K a)$ denotes the $SU(2)$ holonomy matrix of $SU(2)$ connection $a$ along the closed knot $K$.   $R_n = \textrm{Sym}^{\otimes (n-1)} \mathbf{2} \in \textrm{Hom}[SU(2) \rightarrow SU(n)]$ is the $n$-dimensional irreducible representation of $SU(2)$ and we define
\begin{align}
\textrm{tr}_R (g)= \textrm{Tr} ( R (g))\;. \nonumber
\end{align}
Since $\textrm{tr}_R (g) = \textrm{tr}(g^{-1})$, the path-integral does not depend on the orientation choice of $K_I$. To regularize infinity coming from  self-interaction of knots, on the other hand,  one need to introduce framing and the quantum Wilson loop expectation value depends on the choice.  In the above, we include following overall sign factor
\begin{align}
\prod_{I=1}^{\sharp(K)} (-1)^{n_I-1} \label{sign factor}
\end{align}
 which reflects that  $J_{K \bigsqcup	\bigcirc}  = -(q^{1/2} +q^{-1/2}) J_K  $ while an additional Wilson loop along the disjoint unknot $\bigcirc$ in $R=\mathbf{2}$ contribute  a muliplicative factor $(q^{1/2}+q^{-1/2})$ to   the path-integral of Chern-Simons theory \cite{witten1989quantum}.

Wilson loop operator, $\textrm{tr}_R (P \exp (\oint_K a))$, along a knot $K \subset S^3$ with color $R=\textrm{Sym}^{\otimes (n-1)} \mathbf{2}$ in the $SU(2)$  Chern-Simons theory can be alternatively described by monodromy defect in the knot complement $S^3\backslash K$ defined by following boundary condition
\begin{align}
\textrm{boundary condition : }P \exp \left(\oint_{\bigcirc_{\rm m}} a \right) =  
\begin{pmatrix}
 e^{\frac{\pi i n}{k}} & 0
 \\
  0 & e^{-\frac{\pi i n}{k}} 
 \end{pmatrix}\;. \label{b.c. for Jones}
\end{align} 
Using the alternative definition of the Wilson loop operator, the Jones polynomial $J_n (\mathcal{K};q)$ can be given as
\begin{align}
J_{n}\left(\mathcal{K};q= e^{2\pi i/k} \right) 
		=(-1)^{n-1}\frac{ \int \frac{[da]}{(\rm gauge)} \bigg{|}_{\textrm{b.c. in \eqref{b.c. for Jones}}} \exp \left(\frac{i(k-2)}{4\pi} \int_{S^3 \backslash \mathcal{K}} \textrm{CS}(a)\right) }{  \int \frac{[da]}{(\rm gauge)} \exp \left(\frac{i(k-2)}{4\pi} \int_{S^3} \textrm{CS}(a)\right)  }\;. \label{path-integral of jones 2}
\end{align} 

\begin{figure}[htbp]
	\begin{center}
		\includegraphics[width=.25\textwidth]{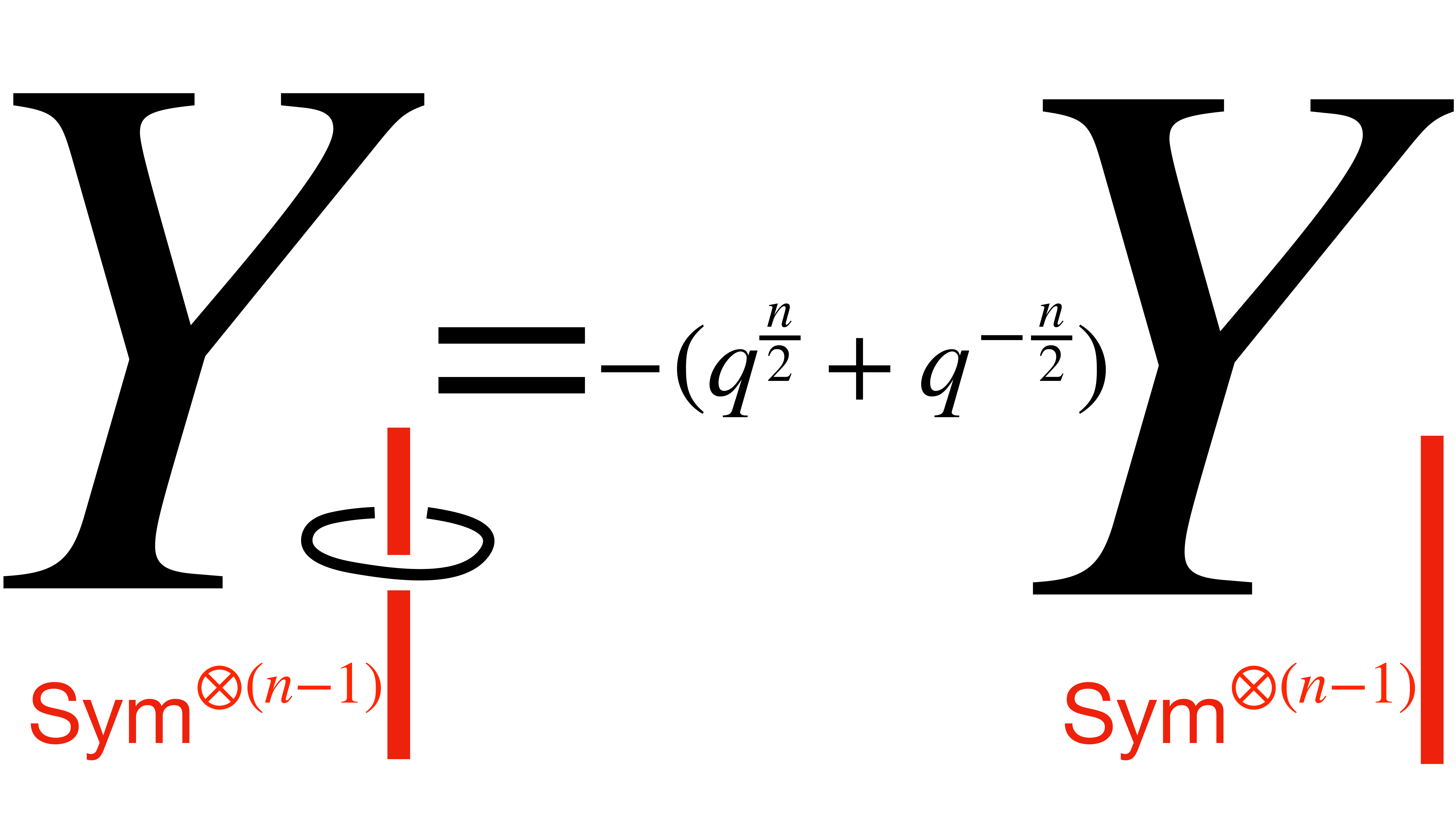}
	\end{center}
	\caption{
	The effect of adding the meridian knot. A  relation between $Y_{({\color{red} \textrm{Sym}^{\otimes (n-1)} \mathcal{K}}) \cup \bigcirc_{\rm m}}$ and $Y_{({\color{red} \textrm{Sym}^{\otimes (n-1)} \mathcal{K}})}$ holds, where  $\bigcirc_{\rm m}$ is a unknot linking the heavy knot $\mathcal{K}$. }
	\label{fig: Reidemeister move 1p}
\end{figure}

Using the  physical definition of the Jones polynomial, now let us explain a non-trivial property of the  knot invariant summarized in Figure \ref{fig: Reidemeister move 1p}.
In the Chern-Simons path-integral,  the addition of the meridian knot $\bigcirc_{\rm m}$ on the top of  the knot $K$ corresponds to including following term in the integrand of \eqref{path-integral of jones 2},
\begin{align}
	(-1)\times \textrm{tr}_{R = \mathbf{2}}  \left(P \exp  ( \oint_{\bigcirc_{\rm m}} a )\right)\;.
\end{align}
Note that the value of the above Wilson loop is always 
\[
-\left(e^{\frac{\pi i n}k }+e^{-\frac{\pi i n}k }\right) =-\left( q^{\frac{n}2}+q^{-\frac{n}2} \right)
\]
for arbitrary gauge field $a$ satisfying the boundary condition in \eqref{b.c. for Jones}. 
It explains the property of the knot invariant depicted in Figure \ref{fig: Reidemeister move 1p}.

\section{Quantum dilogarithm function} \label{App : QDL}

The quantum dilogarithm function (Q.D.L) $\psi_{\hbar} (Z) $ is defined by  \cite{Faddeev:1993rs}
\begin{align}
\begin{split}
\psi_{\hbar}(Z) := 
\begin{cases}
\prod_{r=1}^{\infty} \frac{1 - q^r e^{-Z}}{1 - \tilde{q}^{-r+1} e^{-\tilde{Z} } }  \quad \text{if} \quad |q| < 1\;,
\\
\prod_{r=1}^{\infty} \frac{1 - \tilde{q}^r e^{-\tilde{Z}}}{1 - q^{-r+1} e^{-Z } }  \quad \text{if} \quad |q| > 1\;,
\end{cases}
\end{split}
\end{align}
with
\begin{align}
\begin{split}
q := e^{\hbar}
\,,\qquad
\tilde{q} := e^{-\frac{4\pi^2}{\hbar}}
\,,\qquad
\tilde{Z} := \frac{2\pi i Z}{\hbar }.
\end{split}
\end{align}
%
The function satisfies the following difference equations:
\begin{align}
\begin{split}
\psi_{\hbar}(Z+\hbar) = (1-e^{-Z}) \psi_{\hbar}(Z)
\,,\quad
\psi_{\hbar}(Z+2\pi i ) = (1-e^{-\tilde{Z}}) \psi_{\hbar}(Z)\;. \label{difference for QDL}
\end{split}
\end{align}
In the asymptotic limit $\hbar \rightarrow 0$,
\begin{align}
\begin{split}
\log \psi_{\hbar}(Z) \xrightarrow{\hbar\rightarrow 0}
\sum_{n=0}^{\infty} \frac{B_n \hbar^{n-1}}{n !} \text{Li}_{2-n}(e^{-Z})\;.
\label{eq:psi expansoin}
\end{split}
\end{align}
Here $B_n$ is the $n$-th Bernoulli number with $B_1=1/2$. 
The Q.D.L satisfies the following identity
\begin{align}
\int \frac{dZ}{\sqrt{2\pi \hbar}} e^{-\frac{Z U}\hbar}\psi_\hbar (Z) = e^{i \delta}e^{\frac{U^2-(2\pi i +\hbar)U }{2\hbar} +\frac{i \pi (b^2+b^{-2})}{12 }} \psi_\hbar (U) \label{Fourier of QDL}
\end{align}
with a constant overall phase factor $e^{i\delta}$.

\newpage



\bibliographystyle{abbrv}
\bibliography{CMP-draft}

\end{document}